\documentclass[osajnl,twocolumn,showpacs,superscriptaddress,10pt]{revtex4-1}
\usepackage{amsmath,amssymb,graphicx}
\begin{document}
\title{Electromagnetic Wave Transmission Through a Subwavelength
 Nano-hole in a Two-dimensional Plasmonic Layer}

\author{Norman J. M. Horing}
\email{Corresponding author: nhoring@stevens.edu}
\author{D\'{e}sir\'{e} Miessein}
\affiliation{Department of Physics and Engineering Physics,
Stevens Institute of Technology, Hoboken, New Jersey 07030, USA}

\author{Godfrey Gumbs}
\affiliation{Department of Physics and Astronomy,
Hunter College of the City University of New York,
695 Park Avenue, New York, NY 10065, USA}

\date{\today}

\begin{abstract}
An integral equation is formulated to describe electromagnetic wave transmission through
a subwavelength nano-hole in a thin plasmonic sheet in terms of the dyadic Green's
function for the associated Helmholtz problem. Taking the subwavelength radius of the
nano-hole to be the smallest length of the system, we have obtained an exact solution
of the integral equation for the dyadic Green's function analytically and in closed form.
This dyadic Green's function is then employed in the numerical analysis of electromagnetic wave
transmission through the nano-hole for normal incidence of the incoming wave train.
The electromagnetic transmission involves two distinct contributions, one emanating from
the nano-hole and the other is directly transmitted through the thin plasmonic layer itself
(which would not occur in the case of a perfect metal screen).  The transmitted radiation
exhibits interference fringes in the vicinity of the nano-hole, and they tend to flatten as
a function of increasing lateral separation from the hole, reaching the uniform
value of transmission through the sheet alone at large separations.
\end{abstract}


\maketitle
\def\thesection{\arabic{section}}
\def\thesubsection{\arabic{subsection}}
\newpage

\section{Introduction}
The description of electromagnetic wave transmission through an aperture in a screen involves
the solution of Maxwell's equations with complicated boundary conditions that are often not really known(or effectively implemented) for the materials
involved.  The diffraction of light by such subwavelength apertures (radius $ R \ll \lambda$) has attracted a great deal of
interest by many researchers:  In an elegant analysis 70 years ago, Bethe [1] showed that the classic Kirchoff method is probably
approximately valid only for large apertures, but not for small ones, and Bethe developed a theoretical approach appropriate
to the subwavelength regime.  Important further developments involving variational principles and dyadic Green's functions were
presented by Levine and Schwinger [2], [3].
However, all of the works we have cited above assume the screen to be a perfect metallic conductor,
whereas intense current interest is focussed on plasmonic semiconductors.  To address such systems and incorporate the issue of
boundary conditions we employ an integral equation formulation [4],[5],[6] describing electromagnetic wave
transmission through a nano-hole in a thin plasmonic layer in terms of the dyadic Green's function
for the associated vector Helmholtz problem [7],[8],[9],[10].
Interesting electromagnetic phenomena, including enhanced transmission through a nano-hole in an opaque screen, have been discussed in
the literature mostly for a metallic thick screen [11],[12],[13],[14].
The aim of the present work is to study electromagnetic wave field transmission and reflection at a \textit{thin} semiconductor screen including the
role of the embedded 2D plasmonic layer, and transmission directly through it as well as through the aperture, thereby improving our understanding
of subwavelength electromagnetic phenomenology.

This paper is structured as follows: in Section $ 2 $, we discuss the dyadic Green's function
solution for a thin (2D) semiconductor plasma layer located at the plane $ z = 0 $ in a 3D bulk host medium with background dielectric constant $\varepsilon_{b}^{(3D)}$.
Section $ 3 $ deals with the description of a nano-hole in the dyadic Green's function-integral equation for the 2D plasmonic layer following the techniques of references [4] and [5].
In Section $ 4 $, we address the electromagnetic wave transmission of a perforated thin screen.  These results are then used in Section $ 5 $ to obtain the transmitted/reflected
electromagnetic field in terms of an infinite incident plane wave train.
Calculated results for normal incidence are exhibited in Section $ 6 $ for the near-field ($ z=50 R $), for intermediated-field ($ z=250 R $) and
for far-field ($ z = 1000 R $) zones of diffraction.

\section{ Dyadic Electromagnetic Green's Function for a Thin Semiconductor Layer }

\begin{figure}[h]
\centering
\includegraphics[width=5cm,height=4cm]{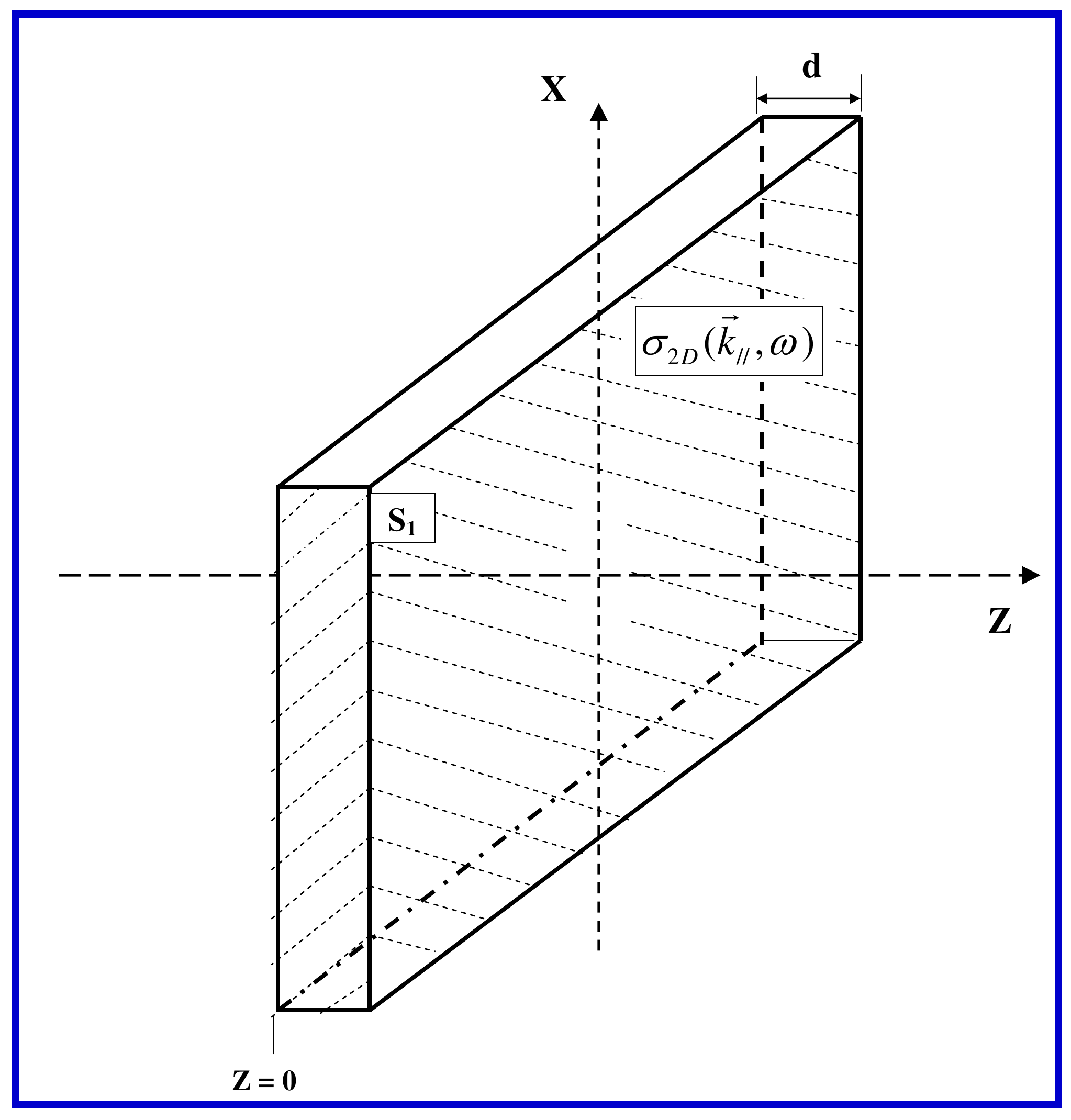}
\caption{Two dimensional plasmonic layer of thickness $ d \ll \lambda $ (small compared to the wavelength $ \lambda $) embedded at $ z=0 $ in a
three dimensional bulk host medium with dielectric constant $ \varepsilon_{b}^{(3D)}$.}
\label{FIG1HL}
\qquad
\end{figure}

The description of electrodynamics in terms of dyadic Green's functions
has a long history in physics and in electrical engineering [7],  [8], [9], [10].
The present interest in the electrodynamics of low-dimensional semiconductor nanostructures,
which involve currents that are geometrically confined in narrow regions, offers a fertile ground
for the application of dyadic Green's functions.  To deal with such problems we
introduce the description of electromagnetic response of various semiconductor
nanostructures in terms of a dyadic
Green's function propagator [2], [8], [9].
To start, we focus on a two dimensional plasmonic layer $ S_{1}$ located on the plane
$ z = 0 $, embedded in a three dimensional bulk host medium with
background dielectric constant $\varepsilon_{b}^{(3D)}$ (Fig.\ref{FIG1HL}), and will examine the electromagnetic wave transmission
through it using the dyadic Green's function method.
The Helmholtz equation for the electric field, based on the Maxwell equations, is given in $\omega-$ frequency representation as
\begin{equation}\label{A2.1}
\left[ \widehat{I}\left(\vec{\nabla}^{2}+\frac{\,\omega^{2}}{c^{2}}\right)-
\vec{\nabla}\vec{\nabla}\right]\cdot\vec{E}(\vec{r},\omega) = -\,\frac{4\,\pi\,i\,\omega}{c^{2}}\,\vec{J}(\vec{r},\omega),
\end{equation}
where $ \widehat{I}=\widehat{x}\widehat{x}+\widehat{y}\widehat{y}+\widehat{z}\widehat{z}$  is a unit dyadic tensor and $ c $ is the speed of light in vacuum.
This field is driven by the total electric current density $ \vec{J}(\vec{r},t)$ at space-point $ \vec{r}$ at time $ t $, which includes the induced current density,
$ \vec{J}_{ind}(\vec{r},t)$ (taken to be linear) composed of $ (i) $ a term arising from the conductivity of the local bulk host medium,
\begin{equation}\label{A2.2}
    \widehat{\sigma}^{(3D)}_{b}(\vec{r},\vec{r}^{'};\omega)=\widehat{I}\,\frac{i\,\omega}{4\,\pi}
\left[1-\varepsilon_{b}^{(3D)}\right]\delta^{(3D)}(\vec{r}-\vec{r}^{'})
\end{equation}
($ \delta^{(3D)}(\vec{r}-\vec{r}^{'})$ is the 3D Dirac delta function) and
$ (ii) $  a term arising from the conductivity of the thin plasmonic layer $ \widehat{\sigma}^{(2D)}_{fs}(\vec{r},\vec{r}^{'};\omega)$
under consideration.
Including the role of the external driving current density $ \vec{J}_{ext}(\vec{r},t)$, the electromagnetic Helmholtz equation can then be written as
\begin{eqnarray}\label{A2.3}
\left[\widehat{I}\left(\vec{\nabla}^{2} \,+ \,
\frac{\omega^{2}}{c^{2}}\varepsilon_{b}^{(3D)}\right)
- \vec{\nabla}\vec{\nabla}\right]\vec{E}(\vec{r},\omega)
\nonumber\\
\,+\,\frac{4\pi\,i\omega}{c^{2}}
\int d^{3}\vec{r}^{'}{\widehat{\sigma}_{fs}}^{(2D)}(\vec{r},\vec{r}^{'};\omega)
\vec{E}(\vec{r}^{'};\omega)
\nonumber\\
=-\frac{4\pi\,i\omega}{c^{2}}\vec{J}_{ext}(\vec{r};\omega).
\end{eqnarray}

In general, in electromagnetic theory for linear materials, the dyadic Green's function $ \widehat{G} $ is defined as the
electric field $ \vec{E}$ at the field point $ \vec{r}$ generated by a radiating unit dyadic impulsive source located at the point $ {\vec{r}}^{\,\prime}$,
so that, the dyadic Green's function relates $ \vec{E}(\vec{r},\omega) $ to the current density $ \vec{J}(\vec{r}^{'},\omega) $ as [3], [4], [7]
\begin{equation}\label{A2.4}
\vec{E}(\vec{r},\omega)=\frac{4\pi\,i\omega}{c^{2}}\int d^{3}\vec{r}\,\,{\widehat{G}}(\vec{r},\vec{r}^{'};\omega)\cdot\vec{J}_{ext}(\vec{r}^{'};\omega).
\end{equation}
To fulfill the role of $ {\widehat{G}}(\vec{r},\vec{r}^{'};\omega)$ mandated by  Eq.(\ref{A2.4}) with  Eq.(\ref{A2.3}), we define
the dyadic Green's function of the homogeneous layer, $ {\widehat{G}}(\vec{r},\vec{r}^{'};\omega)$ $\longrightarrow $
$ {\widehat{G}_{fs}}(\vec{r},\vec{r}^{'};\omega)$ as
\begin{eqnarray}\label{A2.5}
\left[\widehat{I}\left(\vec{\nabla}^{2}\,+\,
\frac{\omega^{2}}{c^{2}}\varepsilon_{b}^{(3D)}\right)
- \vec{\nabla}\vec{\nabla}\right]{\widehat{G}_{fs}}(\vec{r},\vec{r}^{'};\omega)
\nonumber\\
+\,\frac{4\pi\,i\omega}{c^{2}}
\int d^{3}\vec{r}^{''}{\widehat{\sigma}_{fs}}^{(2D)}(\vec{r},\vec{r}^{''};\omega)
{\widehat{G}_{fs}}(\vec{r}^{''},\vec{r}^{'};\omega)
\nonumber\\
=\,-\,\widehat{I}{\delta^{(3D)}}(\vec{r}-\vec{r}^{'}).
\end{eqnarray}
Following the position-space inversion of the differential operator in the brackets on the Left Hand Side,
carried out in the analysis of Refs. [4], [5] and [6], Eq.(\ref{A2.5}) may be written in the form of a full integral equation as
\begin{eqnarray}\label{A2.6}
{\widehat{G}_{fs}}(\vec{r},\vec{r}^{'};\omega)&=&
{\widehat{G}_{3D}}(\vec{r},\vec{r}^{'};\omega)
\nonumber\\
\,&+&\,
\frac{4\pi\,i\omega}{c^{2}}\int d^{3}\vec{r}^{''}
\int d^{3}\vec{r}^{'''}{\widehat{G}_{3D}}(\vec{r},\vec{r}^{''};\omega)
\nonumber\\
&\times&\,
{\widehat{\sigma}_{fs}}^{(2D)}(\vec{r}^{"},\vec{r}^{'''};\omega)
{\widehat{G}_{fs}}(\vec{r}^{'''},\vec{r}^{'};\omega).
\nonumber\\
\end{eqnarray}
Because of spatial translational invariance in the plane of the plasmonic layer,
the dyadic Green's function may be written in terms of a Fourier transform in the parallel plane
\newline
$[\vec{r}_{\parallel}-{\vec{r}_{\parallel}}^{\,'}\mapsto\vec{k}_{\parallel}]$,
so that Eq.(\ref{A2.6}) takes the form
\begin{eqnarray}\label{A2.7}
{\widehat{G}_{fs}}({\vec{k}}_{\parallel};z,{z}^{'};\omega)&=&
{\widehat{G}_{3D}}({\vec{k}}_{\parallel};z,{z}^{'};\omega)
\nonumber\\
&+&\,
\frac{4\pi\,i\omega}{c^{2}}\int d{z}^{''}
\int d{z}^{'''}{\widehat{G}_{3D}}({\vec{k}}_{\parallel};z,{z}^{''};\omega)
\nonumber\\
&\times&\,
{\widehat{\sigma}_{fs}}^{(2D)}({\vec{k}}_{\parallel};{z}^{''},{z}^{'''};\omega)
{\widehat{G}_{fs}}({\vec{k}}_{\parallel};{z}^{'''},{z}^{'};\omega),
\nonumber\\
\end{eqnarray}
where $\vec{k}_{\parallel}=k_{x}\widehat{e}_{x}+k_{y}\widehat{e}_{y}$.
In this matter, the response of the uniform two dimensional plasmonic layer is described by the local conductivity tensor
$ \widehat{\sigma}_{fs}^{(2D)}$ as
\begin{eqnarray}\label{A2.8}
{\widehat{\sigma}_{fs}}^{(2D)}({\vec{k}}_{\parallel};{z}^{''},{z}^{'''};\omega)
={\widehat{I}}\,{\sigma}_{fs}^{(2D)}({\vec{k}}_{\parallel};\omega)\delta(z^{''})\delta(z^{'''})
\end{eqnarray}
where
\begin{equation}\label{A2.9}
{\sigma}_{fs}^{(2D)}({\vec{k}}_{\parallel};\omega)
=\frac{i\,\omega}{4\pi}\left[ \varepsilon_{b}^{(3D)}-\varepsilon(\omega)\right]d
\end{equation}
and $ \varepsilon(\omega)$ represents the dielectric function of the 2D plasmonic layer of thickness $ d $.

With this, the solution of Eq.(\ref{A2.7}) for ${\widehat{G}_{fs}}({\vec{k}}_{\parallel};z,{z}^{'};\omega)$ was determined in Refs.  [4],[5] and [6] as
\begin{eqnarray}\label{A2.10}
{\widehat{G}_{fs}}({\vec{k}}_{\parallel};z,{z}^{'};\omega) &=&
{\widehat{G}_{3D}}({\vec{k}}_{\parallel};z,{z}^{'};\omega)
\nonumber\\
&+&
\gamma\,
{\widehat{G}_{3D}}({\vec{k}}_{\parallel};z,0;\omega)\,
\nonumber\\
&\times &
\left[\widehat{I}\,-\,\gamma\,
{\widehat{G}_{3D}}({\vec{k}}_{\parallel};0,0;\omega)\, \right]^{-1}
\nonumber\\
&\times &
{\widehat{G}_{3D}}({\vec{k}}_{\parallel};0,{z}^{'};\omega)
\end{eqnarray}
with
\begin{equation}\label{A2.11}
    \gamma\,=\,\frac{4\pi\,i\,\omega}{c^{2}}\,{\sigma}_{fs}^{(2D)}({\vec{k}}_{\parallel};\omega)
\end{equation}
and
\tiny

\begin{eqnarray}\label{A2.12}
{\widehat{G}_{3D}}({\vec{k}}_{\parallel};z,{z}^{'};\omega) = \,-\,
\frac{e^{i\,k_{z}\mid z-z^{'}\mid}}{2\,i\,k_{z}} &
\nonumber\\
\times
\left\{\widehat{I}-\frac{1}{q_{\omega}^{2}}\left[\vec{k}_{\parallel}\,\vec{k}_{\parallel}+
k_{z}\,sgn(z-z^{'})\left(\vec{k}_{\parallel}\widehat{e}_{z}+\widehat{e}_{z}\vec{k}_{\parallel}\right)
+
\widehat{e}_{z}\widehat{e}_{z}\left(k_{z}^{2}-2 i k_{z}\delta(z-z^{'})\right)\right]\right\},
\nonumber\\
\end{eqnarray}

\normalsize
in mixed ($ \vec{k}_{\parallel}; z,z^{'} $) Fourier representation with $ q_{\omega}=(\omega/c)\sqrt{\varepsilon_{b}^{(3D)}} $.
Furthermore,
\begin{equation}\label{A2.13}
    k_{z}= \sqrt{q_{\omega}^{2}-k_{\parallel}^{2}}
\end{equation}
and
\begin{equation}\label{A2.14}
sgn(\overline{z})= \left\{
         \begin{array}{ll}
               1 , & \hbox{ for $ \overline{z} $ $ > 0 $ ;} \\
              0 , & \hbox{ for $ \overline{z} $ $ = 0 $ ;} \\
               -1 , & \hbox{ for $ \overline{z} $ $ < 0 $  ,}
          \end{array}
        \right.
\end{equation}
where $ \overline{z}\,=\, z-z^{'} $.
The sign before the radical has been chosen in such a way that the field in the host
medium satisfies the radiation condition for $ k_{\parallel} < q_{\omega}$ when the radical is purely real, whereas
it represents an evanescent field for $ k_{\parallel}>q_{\omega}$ when the radical is purely imaginary.

The closed form for $ {\widehat{G}_{fs}}({\vec{k}}_{\parallel};z,{z}^{'};\omega)$ of Eq.(\ref{A2.10})
requires evaluation of
\newline
$ {\widehat{G}_{3D}}({\vec{k}}_{\parallel};z,{z}^{'};\omega)$ at $ z = 0 $ and $ z^{'} = 0 $
\small
\begin{eqnarray}\label{A2.15}
{\widehat{G}_{3D}}({\vec{k}}_{\parallel};0,0;\omega) &=&\,-\,
\frac{1}{2\,i\,k_{z}}
\nonumber\\
& \times &
\left\{\widehat{I}-\frac{1}{q_{\omega}^{2}}\left[\vec{k}_{\parallel}\,\vec{k}_{\parallel}+
\widehat{e}_{z}\widehat{e}_{z}\left(k_{z}^{2}-\frac{2 i k_{z}}{d})\right)\right]\right\},
\nonumber\\
\end{eqnarray}
\normalsize
where the Dirac delta function evaluated at the origin ($ z=0$)$, \delta(0)$, is approximated by the inverse of
the plasmonic layer thickness, $\frac{1}{d}$, and $ sgn(0) = 0 $.  This approximation of $ {\widehat{G}_{3D}}$ is undertaken
in the context of the integral equation for $ {\widehat{G}_{fs}}$, Eq.(\ref{A2.7}) in which $ {\widehat{G}_{3D}}$ is involved in the kernel.
In this regard, the zero width of the $ z^{''}$-integration that is mandated by Eq.(\ref{A2.8}), is recognized as artificial, and is more realistically
smeared over a small (but finite) range $ \Delta z^{''} $ $\sim d $.

The determination of ${\widehat{G}_{fs}}({\vec{k}}_{\parallel};z,0;\omega)$ requires the 3D matrix inversion of
\newline
$ \left[\widehat{I}\,-\,\gamma\,{\widehat{G}_{3D}}({\vec{k}}_{\parallel};0,0;\omega) \right] $ and, specifically,
$ {\widehat{G}_{fs}}({\vec{k}}_{\parallel};z,0;\omega) $ is found to be:
\small
\begin{eqnarray}\label{A2.16}
{\widehat{G}_{fs}}({\vec{k}}_{\parallel};z,0;\omega) = {\widehat{G}_{3D}}({\vec{k}}_{\parallel};z,0;\omega)\,\left[\widehat{I}\,- \,\gamma\,
{\widehat{G}_{3D}}({\vec{k}}_{\parallel};0,0;\omega) \right]^{-1}.
\nonumber\\
\end{eqnarray}
\normalsize
Denoting the dyad to be inverted as $ \widehat{\Omega}$
\tiny
\begin{eqnarray}\label{A2.17}
\widehat{\Omega}&=&\left[\widehat{I}\,-\, \,\gamma\,
{\widehat{G}_{3D}}({\vec{k}}_{\parallel};0,0;\omega) \right]
\nonumber\\
&=&
\begin{bmatrix}
  1\,-\,\gamma\,G_{3D}^{xx}({\vec{k}}_{\parallel};0,0;\omega) & \,-\,\gamma\,G_{3D}^{xy}({\vec{k}}_{\parallel};0,0;\omega) & 0 \\
  \,-\,\gamma\,G_{3D}^{yx}({\vec{k}}_{\parallel};0,0;\omega) & 1\,-\,\gamma\,G_{3D}^{yy}({\vec{k}}_{\parallel};0,0;\omega) & 0 \\
  0 &0 & 1\,-\,\gamma\,G_{3D}^{zz}({\vec{k}}_{\parallel};0,0;\omega)
 \end{bmatrix}.
 \nonumber\\
\end{eqnarray}
\normalsize
Noting that $ \widehat{\Omega}$ is block diagonal, its  2$\times$2 block is readily inverted and its $ zz $-element inverts algebraically.
The matrix elements of $ \widehat{\Omega}^{-1}$ are written out explicitly in Appendix A.

Forming the product
\begin{equation}\label{A2.18}
   {\widehat{G}_{fs}}({\vec{k}}_{\parallel};z,0;\omega) = {\widehat{G}_{3D}}({\vec{k}}_{\parallel};z,0;\omega)\,\,\widehat{\Omega}^{-1}
\end{equation}
using the matrix elements provided in Appendix A, we employ a coordinate system in which $ k_{y}=0$, with the result(see Appendix C):
\tiny
\begin{eqnarray}\label{A2.19}
  {\widehat{{G}}_{fs}}({k}_{x},{k}_{y}=0;z,0;\omega) =
  \nonumber\\
  \begin{bmatrix}
  {G}_{fs}^{xx}({k}_{x},{k}_{y}=0;z,0;\omega) & 0 & {G}_{fs}^{xz}({k}_{x},{k}_{y}=0;z,0;\omega) \\
  0 & {G}_{fs}^{yy}({k}_{x},{k}_{y}=0;z,0;\omega) & 0 \\
  {G}_{fs}^{zx}({k}_{x},{k}_{y}=0;z,0;\omega) & 0 & {G}_{fs}^{zz}({k}_{x},{k}_{y}=0;z,0;\omega)
 \end{bmatrix}
 \nonumber\\
\end{eqnarray}
\normalsize
where
\small
\begin{equation}\label{A2.20}
{G}_{fs}^{xx}({k}_{x},{k}_{y}=0;z,0;\omega)=\,-\,\frac{e^{i\,k_{z}\mid z\mid }}{2\,i\,k_{z}}\left[\frac{1}{\overline{D}_{1}}\left\{\left(1-\frac{k_{x}^{2}}{q_{\omega}^{2}}\right) \right\}\right],
\end{equation}
\begin{equation}\label{A2.21}
{G}_{fs}^{yy}({k}_{x},{k}_{y}=0;z,0;\omega)=\,-\,\frac{e^{i\,k_{z}\mid z\mid }}{2\,i\,k_{z}}\left[\frac{1}{\overline{D}_{3}}\right],
\end{equation}
\begin{eqnarray}\label{A2.22}
{G}_{fs}^{zz}({k}_{x},{k}_{y}=0;z,0;\omega)&=&\,-\,\frac{e^{i\,k_{z}\mid z\mid }}{2\,i\,k_{z}}
\nonumber\\
&\times &
\left[\frac{1}{\overline{D}_{2}}\left\{\left(1-\frac{k_{z}^{2}- 2 i k_{z} \delta (z)}{q_{\omega}^{2}}\right)\right\}\right],
\nonumber\\
\end{eqnarray}
\begin{equation}\label{A2.23}
{G}_{fs}^{xy}({k}_{x},{k}_{y}=0;z,0;\omega)=0,
\end{equation}
\begin{equation}\label{A2.24}
{G}_{fs}^{xz}({k}_{x},{k}_{y}=0;z,0;\omega)=\,+\,\frac{e^{i\,k_{z}\mid z\mid }}{2\,i\,k_{z}}\left[\frac{1}{\overline{D}_{2}}\left(\frac{k_{x}\,k_{z}sgn(z)}{q_{\omega}^{2}}\right)\right],
\end{equation}
\begin{equation}\label{A2.25}
{G}_{fs}^{yz}({k}_{x},{k}_{y}=0;z,0;\omega)=0,
\end{equation}
\begin{equation}\label{A2.26}
{G}_{fs}^{yx}({k}_{x},{k}_{y}=0;z,0;\omega) = 0,
\end{equation}
\begin{equation}\label{A2.27}
{G}_{fs}^{zx}({k}_{x},{k}_{y}=0;z,0;\omega)=\,+\,\frac{e^{i\,k_{z}\mid z\mid }}{2\,i\,k_{z}}\left[\frac{1}{\overline{D}_{1}}\left(\frac{k_{z}\,k_{x}sgn(z)}{q_{\omega}^{2}}\right)\right],
\end{equation}
\begin{equation}\label{A2.28}
{G}_{fs}^{zy}({k}_{x},{k}_{y}=0;z,0;\omega)=0,
\end{equation}
\normalsize
and ( note that $ \overline{D}_{1,2,3}$ defined here is not to be confused with $ {D}_{1,2,3}$ in Appendix C)
\begin{equation}\label{A2.29}
   \overline{D}_{1}=\left[1+ \left(\frac{\gamma}{2\,i\,k_{z}}\right) \left( 1- \frac{k_{x}^{2}}{q_{\omega}^{2}}\right) \right],
\end{equation}
\begin{equation}\label{A2.30}
   \overline{D}_{2}=\left[1+ \left(\frac{\gamma}{2\,i\,k_{z}}\right) \left( 1- \frac{a_{0}^{2}}{q_{\omega}^{2}}\right) \right],
\end{equation}
\begin{equation}\label{A2.31}
   \overline{D}_{3}=\left[1+ \left(\frac{\gamma}{2\,i\,k_{z}}\right)\right]
\end{equation}
with $ a_{0}^{2}=k_{z}^{2}\,-\,\frac{2\,i\,k_{z}}{d} $.
It is interesting to note that Eq.(\ref{A2.19}) can be rewritten as a sum
of diagonal and anti-diagonal dyads as follows
\tiny
\begin{eqnarray}\label{A2.32}
  {\widehat{{G}}_{fs}}({k}_{x},{k}_{y}=0;z,0;\omega)=
  \nonumber\\
  \begin{bmatrix}
  {G}_{fs}^{xx}({k}_{x},{k}_{y}=0;z,0;\omega) & 0 & 0 \\
  0 & \frac{1}{2}{G}_{fs}^{yy}({k}_{x},{k}_{y}=0;z,0;\omega) & 0 \\
  0 & 0 & {G}_{fs}^{zz}({k}_{x},{k}_{y}=0;z,0;\omega)
 \end{bmatrix}
 \nonumber\\
 +
 \newline
 \begin{bmatrix}
 0 & 0 & {G}_{fs}^{xz}({k}_{x},{k}_{y}=0;z,0;\omega) \\
  0 &  \frac{1}{2}{G}_{fs}^{yy}({k}_{x},{k}_{y}=0;z,0;\omega) & 0 \\
 {G}_{fs}^{zx}({k}_{x},{k}_{y}=0;z,0;\omega) & 0 & 0
 \end{bmatrix}.
\nonumber\\
\end{eqnarray}
\normalsize

The dispersion relations for electromagnetic wave modes for the full \textit{non-perforated} plasmonic sheet may be obtained by examining the vanishing determinant of the corresponding dyadic Green's function ${\widehat{{G}}_{fs}}({k}_{x},{k}_{y}=0;z,0;\omega)$.  These normal mode frequencies defined by the vanishing denominators $ \overline{D}_{1}, \overline{D}_{2}, \overline{D}_{3}\mapsto 0 $ are given by
\begin{eqnarray}\label{A2.33}
   \overline{D}_{1}=1\,+\,\frac{2\,\pi\,\omega\, \sigma_{fs}^{(2D)}(\vec{k}_{\parallel},\omega)\,k_{z}}{c^{2}\,q_{\omega}^{2}}=0,
\end{eqnarray}
\begin{eqnarray}\label{A2.34}
   \overline{D}_{2}&=&\,1\,+\,\frac{2\,\pi\,\omega\, \sigma_{fs}^{(2D)}(\vec{k}_{\parallel},\omega)}{c^{2}\,k_{z}}
   \nonumber\\
&\times &
   \left[1\,
   -\,
   \frac{1}{q_{\omega}^{2}}\left(k_{z}^{2}-\frac{2\,i\,k_{z}}{d}\right)\right]=0,
   \nonumber\\
\end{eqnarray}
\begin{eqnarray}\label{A2.35}
   \overline{D}_{3}=\,1\,+\,\frac{2\,\pi\,\omega\, \sigma_{fs}^{(2D)}(\vec{k}_{\parallel},\omega)}{c^{2}\,k_{z}}=0.
\end{eqnarray}

\section{Nano-hole in the Dyadic Green's Function Integral Equation of a 2D Plasmonic Layer }
\begin{figure}[h]
\centering
\includegraphics[width=6cm,height=5cm]{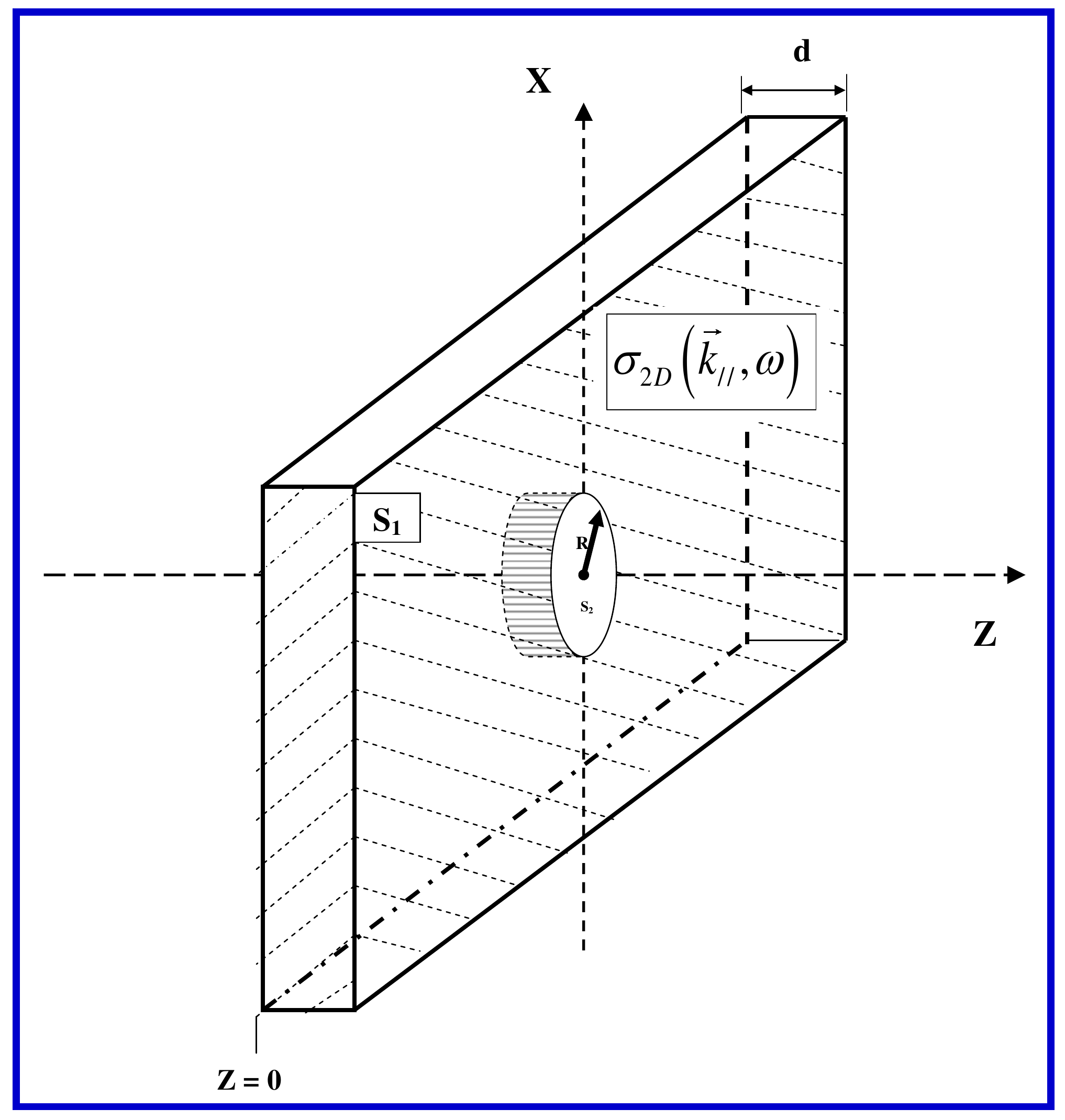}
\caption{Two dimensional plasmonic layer (thickness $ d $, embedded
at $ z=0 $ in a three dimensional bulk meduim) with a nano-hole of area $ A $ at the origin of the $(x-y)$ plane.}
\label{FIG1IHL}
\qquad
\end{figure}

 We consider a two dimensional plasmonic layer $ S $  perforated by a nano-scale aperture of area $ A $ in the $(x-y)$ plane,
representing the nano-hole by subtracting from $ \widehat{\sigma}_{fs}^{(2D)}$ the part of the full sheet conductivity associated with
the hole $ \widehat{\sigma}_{hole}^{(2D)}$ (Fig.\ref{FIG1IHL}),
\begin{eqnarray}\label{A3.1}
\widehat{\sigma}^{(2D)}(\vec{r},\vec{r}^{'};\omega)\,=\,\widehat{\sigma}_{fs}^{(2D)}(\vec{r},\vec{r}^{'};\omega)
\,-\,\widehat{\sigma}_{hole}^{(2D)}(\vec{r},\vec{r}^{'};\omega).
\end{eqnarray}
Accordingly, the dyadic Green's function for the perforated screen satisfies the integral equation
in position-frequency representation given by:
\begin{eqnarray}\label{A3.2}
\widehat{G}(\vec{r},\vec{r}^{'};\omega)&=&\widehat{G}_{fs}(\vec{r},\vec{r}^{'};\omega)
\nonumber\\
&\,-\,&
\frac{4\pi\,i\omega}{c^2}\int{d^{3}\vec{r}^{''}}\int{d^{3}\vec{r}^{'''}}\widehat{G}_{fs}(\vec{r},\vec{r}^{''};\omega)\,
\nonumber\\
&\times &
\widehat{\sigma}_{hole}^{2D}(\vec{r}^{''},\vec{r}^{'''};\omega)\widehat{G}(\vec{r}^{'''},\vec{r}^{'};\omega)
\end{eqnarray}
where $\widehat{G}_{fs}(\vec{r},\vec{r}^{'};\omega)$ is the dyadic Green's function for the 2D plasmonic layer in the absence of the nano-hole
and $ \widehat{G}_{fs}(\vec{r},\vec{r}^{''};\omega)\,\widehat{\sigma}_{hole}^{2D}(\vec{r}^{''},\vec{r}^{'''};\omega)$ is the kernel of the new integral equation.
The excised part the conductivity defining the hole is described by the localized conductivity tensor
\small
\begin{eqnarray}\label{A3.3}
\widehat{\sigma}^{(2D)}_{hole}(x,x^{'};y,y^{'};z,z^{'};\omega)&=&\widehat{I}\sigma^{(2D)}_{fs}(\omega)
\eta_{+}\left(\frac{a}{2}-|x|\right)\,
\nonumber\\
&\times &
\eta_{+}\left(\frac{a^{'}}{2}-|x^{'}|\right)\,
\eta_{+}\left(\frac{b}{2}-|y|\right)\,
\nonumber\\
&\times &
\eta_{+}\left(\frac{b^{'}}{2}-|y^{'}|\right)\,
\delta(z)\,\delta(z'),
\nonumber\\
\end{eqnarray}
\normalsize
where Eq.(\ref{A2.9})
\begin{equation}\label{A3.4}
     {\sigma}_{fs}^{(2D)}(\omega)=\frac{i\omega}{4\pi}\left[ \varepsilon_{b}^{(3D)}-\varepsilon(\omega)\right]d,
\end{equation}
and $ \eta_{+}(x)$ is the Heaviside unit step function confining the integration range on the 2D sheet to the nano-hole dimensions.
Noting that
\begin{equation}\label{A3.4I}
   \lim_{a\rightarrow 0}\left[\frac{{\eta}_{+}\left(\frac{a}{2}\,-\,\mid x \mid \right)}{a}\right]=\,\delta(x)
\end{equation}
is a representation of the Dirac delta function, we write
\small
\begin{equation}\label{A3.5}
\widehat{\sigma}^{(2D)}_{hole}(\vec{r}_{\parallel},\vec{r}_{\parallel}^{'};z,z^{'};\omega)\approx
\widehat{I} {A}^{2} \sigma_{fs}^{(2D)}(\omega)
\delta^{2D}(\vec{r}_{\parallel})\delta^{2D}(\vec{r}_{\parallel}^{'})\delta(z)\delta(z'),
\end{equation}
\normalsize
where $ A=a\,b=a^{\prime}\,b^{\prime}$ is the area of the nano-hole.
Employing this in Eq.(\ref{A3.2}) to execute all the integrals, we find
\small
\begin{eqnarray}\label{A3.6}
\widehat{G}(\vec{r}_{\parallel},\vec{r}_{\parallel}^{'};z,z^{'};\omega) &=& \widehat{G}_{fs}(\vec{r}_{\parallel},\vec{r}_{\parallel}^{'};z,z^{'};\omega)
\nonumber\\
&-&
\,\beta\,\widehat{G}_{fs}(\vec{r}_{\parallel},0;z,0;\omega)\,
\widehat{G}(0,\vec{r}_{\parallel}^{'};0,z^{'};\omega)
\nonumber\\
\end{eqnarray}
\normalsize
where
\begin{equation}\label{A3.7}
 \beta = \gamma {A}^{2}=\left(\frac{4\,i\pi\omega}{c^2}\sigma^{(2D)}_{fs}(\omega)\right){A}^{2}.
\end{equation}
Setting $ \vec{r}_{\parallel}=0 $ and $ z =0 $ in Eq.(\ref{A3.6}), we obtain $ \widehat{G}(0,\vec{r}_{\parallel}^{'};0,z^{'};\omega)$ as
\begin{eqnarray}\label{A3.8}
\widehat{G}(0,\vec{r}_{\parallel}^{'};0,z^{'};\omega)& = &  {\left[\widehat{I}\,+\,\beta\,\widehat{G}_{fs}(0,0;0,0;\omega)\right]^{-1}}
\nonumber\\
&\times &
\widehat{G}_{fs}(0,\vec{r}_{\parallel}^{'};0,z^{'};\omega),
\end{eqnarray}
so that
\small
\begin{eqnarray}\label{A3.9}
\widehat{G}(\vec{r}_{\parallel},\vec{r}_{\parallel}^{'};z,z^{'};\omega)& = & \widehat{G}_{fs}(\vec{r}_{\parallel},\vec{r}_{\parallel}^{'};z,z^{'};\omega)
\nonumber\\
&-& {\beta \widehat{G}_{fs}(\vec{r}_{\parallel},0;z,0;\omega)}
\nonumber\\
&\times &
\left[{\widehat{I}\,+\,\beta
\widehat{G}_{fs}(0,0;0,0;\omega)}\right]^{-1}
\nonumber\\
&\times &
{\widehat{G}_{fs}(0,\vec{r}_{\parallel}^{'};0,z^{'};\omega)}.
\nonumber\\
\end{eqnarray}
\normalsize
In particular, $ \widehat{G}(\vec{r}_{\parallel},0;z,0;\omega)$ proves to be of special interest, and is given by
\begin{eqnarray}\label{A3.10}
\widehat{G}(\vec{r}_{\parallel},0;z,0;\omega)&=& {\widehat{G}_{fs}(\vec{r}_{\parallel},0;z,0;\omega)}
\nonumber\\
&\times &
\left[ {\widehat{I}\,+\,\beta \widehat{G}_{fs}(0,0;0,0;\omega)}\right]^{-1}.
\end{eqnarray}

It is important to note that $ \widehat{G}_{fs}(0,0;0,0;\omega)$ involves a divergent integral when all its positional arguments vanish.
In terms of lateral wavevector representation
\begin{equation}\label{A3.11}
 \widehat{G}_{fs}(0,0;0,0;\omega) = \frac{1}{2\pi}\int_{0}^{\infty}dk_{\parallel} k_{\parallel}\,\widehat{{G}}_{fs}(\vec{k}_{\parallel};0,0;\omega),
\end{equation}
the divergence may be removed by introducing a cutoff of the $ k_{\parallel}$-integration range, namely that $ k_{\parallel} < 1/R $
for nano-holes of subwavelength dimensions.
Considering the matrix elements of ${\widehat{G}_{fs}}$ (Eq.(\ref{A2.20})-Eq.(\ref{A2.29})) with $ z = 0 $, $ sgn(0)=0 $ and $ \delta(0)\rightarrow 1/d $,
$ {\widehat{G}_{fs}}({\vec{k}}_{\parallel};0,0;\omega)$ is diagonal:
\tiny
\begin{equation}\label{A3.12}
  {\widehat{G}_{fs}}({\vec{k}}_{\parallel};0,0;\omega)=
  \begin{bmatrix}
  {G}_{fs}^{xx}({\vec{k}}_{\parallel};0,0;\omega) & 0 & 0 \\
  0 & {G}_{fs}^{yy}({\vec{k}}_{\parallel};0,0;\omega) & 0 \\
  0 & 0 & {G}_{fs}^{zz}({\vec{k}}_{\parallel};0,0;\omega)
 \end{bmatrix}
\end{equation}
\normalsize
with matrix elements given by
\begin{equation}\label{A3.13}
{G}_{fs}^{xx}({\vec{k}}_{\parallel};0,0;\omega)=\,-\,\frac{1}{2\,i\,k_{z}}\left[\frac{1}{\overline{D}_{1}}\left\{\left(1-\frac{k_{x}^{2}}{q_{\omega}^{2}}\right) \right\}\right],
\end{equation}
\begin{equation}\label{A3.14}
{G}_{fs}^{yy}({\vec{k}}_{\parallel};0,0;\omega)=\,-\,\frac{1}{2\,i\,k_{z}}\left[\frac{1}{\overline{D}_{3}}\right],
\end{equation}
\begin{equation}\label{A3.15}
{G}_{fs}^{zz}({\vec{k}}_{\parallel};0,0;\omega)=\,-\,\frac{1}{2\,i\,k_{z}}\left[\frac{1}{\overline{D}_{2}}\left\{\left(1-\frac{a_{0}^{2}}{q_{\omega}^{2}}\right)\right\}\right],
\end{equation}
and
\begin{equation}\label{A3.16}
   \overline{D}_{1}=\left[1+ \frac{\Gamma}{k_{z}} \left( 1- \frac{k_{x}^{2}}{q_{\omega}^{2}}\right) \right],
\end{equation}
\begin{equation}\label{A3.17}
   \overline{D}_{2}=\left[1+ \frac{\Gamma}{k_{z}} \left( 1- \frac{a_{0}^{2}}{q_{\omega}^{2}}\right) \right],
\end{equation}
\begin{equation}\label{A3.18}
   \overline{D}_{3}=\left[1+ \frac{\Gamma}{k_{z}}\right]
\end{equation}
where
\begin{equation}\label{A3.19}
     \Gamma=\frac{2\pi\omega\,\sigma_{fs}^{(2D)}(\omega)}{c^{2}}.
\end{equation}

The evaluation of the matrix elements of $\widehat{{G}}_{fs}(0,0;0,0;\omega)$
of Eq.(\ref{A3.11}) are evaluated with the $ {k}_{\parallel}\leq 1/R $ - cutoff
as follows  [15]:
\tiny
\begin{eqnarray}\label{A3.20}
{G}_{fs}^{xx}(0,0;0,0;\omega) &=& \,-\, \frac{1}{4\pi\,i\,R}
\left\{
\frac{1}{2\,\Gamma R}+i\,\frac{(q_{\omega} R)^{2}}{(\Gamma R)^{2}}\sqrt{1-(q_{\omega}R)^{2}}-
\frac{(q_{\omega} R)^{3}}{(\Gamma R)^{2}}\right\}
\nonumber\\
 &+& \frac{1}{4\pi\,i\,R}\left\{
\frac{(q_{\omega} R)^{4}}{(\Gamma R)^{3}} \ln\left[ \frac{(q_{\omega} R)^{2}\,+\,i\,R \Gamma \sqrt{1-(q_{\omega}R)^{2}}}
{(q_{\omega} R)^{2}+R \Gamma (q_{\omega}R)}\right] \right\}\,\,;
\nonumber\\
\end{eqnarray}
\begin{eqnarray}\label{A3.21}
{G}_{fs}^{yy}(0,0;0,0;\omega) &=& \,-\,\frac{1}{4\pi\,i R}
\left\{q_{\omega} R\,-\,i\sqrt{1-(q_{\omega}R)^{2}}-\Gamma R \ln\left[ \frac{R \Gamma +\,q_{\omega}\,R}
{ R \Gamma }\right]\right\}
\nonumber\\
&-&\,\frac{1}{4\pi\,i R}
\left\{\Gamma R \ln\left[ \frac{R \Gamma \,+\,i\,\sqrt{1-(q_{\omega}R)^{2}}}
{ R \Gamma }\right]\right\};
\nonumber\\
\end{eqnarray}
\normalsize
and in the case of the $\widehat{z}\,\widehat{z}$-element, Eq.(\ref{A3.11})- Eq.(\ref{A3.15}) yield
\tiny
\begin{eqnarray}\label{A3.22}
{G}_{fs}^{zz}(0,0;0,0;\omega)& =& \,-\,\frac{q_{\omega}^{2}}{4\pi\,i\,\Gamma}
\left\{\int_{0}^{1} du\,u\,\left[\frac{u^{2}-i\,\alpha_{0}\,u-1}{u^{2}-\alpha_{2}\,u-1}\right]\right\}
\nonumber\\
\,&-&\,\frac{q_{\omega}^{2}}{4\pi\,i\,\Gamma}
\left\{\int_{0}^{\frac{1}{\alpha_{3}}\,\sqrt{1-\alpha_{3}^{2}}} dv\,v\,\left[\frac{v^{2}\,-\,\alpha_{0}\,v+1}{v^{2}\,
+\,i\,\alpha_{2}\,v+1}\right]\right\},
\end{eqnarray}
\normalsize
where we have changed the integration variable $ k_{\parallel}\longrightarrow u=iy $ with
$ u^{2}=-v^{2}=\left[1-\left(k_{\parallel}/q_{\omega}\right)^{2}\right] $
and have defined $  \alpha_{0}=\frac{2\,\delta(0)}{q_{\omega}} = \frac{2}{d\,q_{\omega}}$, $  \alpha_{1}=\frac{q_{\omega}}{\Gamma}$, $ \alpha_{2}=\alpha_{1}+i\,\alpha_{0}$
and $ \alpha_{3}=q_{\omega}\,R $. Writing $ I_{1}$ for the first integral and $ I_{2}$ for the second, the integration for $ I_{1}$ is evaluated as
\tiny
\begin{eqnarray}\label{A3.23}
 I_{1}& =& \frac{1}{2}+\frac{\left(-i \alpha _0+\alpha _2\right)}{2\,\sqrt{-4-\alpha _2^2}}
\left\{-4 \,tan^{-1}\left[\frac{-2+\alpha _2}{\sqrt{-4-\alpha _2^2}}\right]+4 \,{tan}^{-1}\left[\frac{\alpha _2}{\sqrt{-4-\alpha _2^2}}\right]\right\}
\nonumber\\
&+&
 \frac{\left(-i \alpha _0+\alpha _2\right)}{\,\sqrt{-4-\alpha _2^2}}
\left\{-{tan}^{-1}\left[\frac{-2+\alpha _2}{\sqrt{-4-\alpha _2^2}}\right]+\,{tan}^{-1}\left[\frac{\alpha _2}{\sqrt{-4-\alpha _2^2}}\right]\right\} \alpha _2^2
\nonumber\\
&+&
\frac{\left(-i \alpha _0+\alpha _2\right)}{\,\sqrt{-4-\alpha _2^2}}
\left\{\sqrt{-4-\alpha _2^2}+\,\frac{1}{2}\,\ln\left[\alpha_{2}\right] \alpha _2 \sqrt{-4-\alpha _2^2}\right\},
\end{eqnarray}
\normalsize
and for $ I_{2}$ we have
\tiny
\begin{eqnarray}\label{A3.24}
 I_{2}& =& tanh^{-1}\left[\frac{\alpha_{2}}{\sqrt{4+\alpha{_2}^2}}\right] \frac{\left(-i \alpha_{0}+\alpha_{2}\right) \left(2\,+\,\alpha_{2}^2\right)}{\sqrt{4\,+\,\alpha_{2}^2}}
\nonumber\\
&+&
\frac{1}{4}\left\{-2\,-\,2\, tan^{-1}\left[\alpha_{2} \alpha_{3} \sqrt{1-\alpha_{3}^2}\right] \left(\alpha_{0}\,+\,i \alpha_{2}\right) \alpha_{2}\,\right\}
\nonumber\\
&+&
\frac{1}{4}\left\{
\,i\, \ln\left[\alpha _2^2 \left(-1+\frac{1}{\alpha _3^2}\right)+\frac{1}{\alpha _3^4}\right] \left(\alpha _0+i \alpha _2\right) \alpha _2\right\}
\nonumber\\
&+&
\frac{1}{4}\left\{4\, tan^{-1}\left[\frac{i \alpha _2 \alpha _3+2 \sqrt{1-\alpha _3^2}}{\sqrt{4+\alpha _2^2} \alpha _3}\right] \frac{\left(\alpha _0\,+\,i \alpha _2\right) \left(2+\alpha _2^2\right)}{\sqrt{4+\alpha _2^2}}\right\}
\nonumber\\
&+&
\frac{1}{4}\left\{
\frac{2}{\alpha _3^2}-\frac{4 \left(\alpha _0+i \alpha _2\right) \sqrt{1-\alpha _3^2}}{\alpha _3}
\right\}
\nonumber\\
\end{eqnarray}
\normalsize
and, finally,
\begin{eqnarray}\label{A3.25}
  {G}_{fs}^{zz}(0,0;0,0;\omega) &=& \, \,-\,\frac{q_{\omega}^{2}}{4\pi\,i\,\Gamma}
\left\{ I_{1}+ I_{2}\right\}
\end{eqnarray}
for $ q_{\omega}R < 1$.
\newpage
Recalling that the final dyadic Green's function $ \widehat{G}(\vec{r}_{\parallel},0;z,0;\omega) $ of the entire system
(Eq.(\ref{A3.10})) involves the inverse of the matrix $ \widehat{M} $ given by $  \widehat{M} = \left[ {\widehat{I}\,+\,\beta\,\widehat{G}_{fs}(0,0;0,0;\omega)}\right]$
where $\widehat{G}_{fs} \equiv \widehat{G}_{fs}(0,0;0,0;\omega)$ is diagonal so that its inverse is given by
\small
\begin{equation}\label{A3.26}
  {\widehat{M}}^{-1} =
 \begin{bmatrix}
 \left({1\,+\,\beta\,G_{fs}^{xx}}\right)^{-1} & 0 & 0 \\
  0 & \left({1\,+\,\beta\,G_{fs}^{yy}}\right)^{-1} & 0 \\
  0 & 0 & \left({1\,+\,\beta\,G_{fs}^{zz}}\right)^{-1}
 \end{bmatrix}
\end{equation}
\normalsize
where
\begin{equation}\label{A3.26I}
 \left\{
         \begin{array}{ll}
            G_{fs}^{xx} = & \hbox{$ G_{fs}^{xx}(0-0;0,0;\omega)$ ;} \\
             G_{fs}^{yy} = & \hbox{$ G_{fs}^{yy}(0-0;0,0;\omega)$ ;} \\
            G_{fs}^{zz} = & \hbox{$ G_{fs}^{zz}(0-0;0,0;\omega)$ }.
         \end{array}
       \right.
\end{equation}
Furthermore, Eq.(\ref{A3.10}) requires evaluation of $ \widehat{G}_{fs} $ in the lateral position representation
(as opposed to $ k_{\parallel}$-representation) for the full 2D homogeneous plasmonic layer in the absence of the nano-hole.
Correspondingly, the spatial Fourier transform of $ \widehat{{G}}_{fs}(\vec{k}_{\parallel},0;z,0;\omega)$,
\small
\begin{eqnarray}\label{A3.27}
 \widehat{G}_{fs}(\vec{r}_{\parallel},0;z,0;\omega)&=&
 \frac{1}{(2\pi)}\int_{0}^{\infty} dk_{\parallel} k_{\parallel}\,J_{0}({k}_{\parallel}{r}_{\parallel}\,)
\widehat{{G}}_{fs}(\vec{k}_{\parallel};z,0;\omega).
\nonumber\\
\end{eqnarray}
\normalsize
As the convergent integrals involved here ($ z\neq 0 $) are difficult to carry
out analytically, we evaluate them numerically using Eq.(\ref{A2.19}) - Eq.(\ref{A2.28})
in the $ k_{\parallel}$-integrands as follows ($ \,k_{z}=\sqrt{q_{\omega}^{2}\,
-\,k_{\parallel}^{2}}\,$ and set $ k_{\parallel}\equiv q_{\omega}y ,
 \rho=q_{\omega}r_{\parallel}$ and $|s|=q_{\omega}|z|$):
\tiny
\begin{eqnarray}\label{A3.28}
{G}_{fs}^{xx}(\vec{r}_{\parallel},0;z,0;\omega)&=& -\frac{1}{4\pi\,i}\frac{q_{\omega}^{2}}{\Gamma}\left\{\int_{0}^{1}{d}{y}\,y\,J_{0}(\rho y)\,
e^{i\,\mid s \mid \sqrt{1-y^{2}}}\right\}
\nonumber\\
&-&\frac{1}{4\pi\,i}\frac{q_{\omega}^{2}}{\Gamma}\left\{
\int_{1}^{\infty}{d}{y}\,y\,J_{0}(\rho y)\,
e^{-\mid s \mid \sqrt{y^{2}-1}}\right\}
\nonumber\\
&+& \frac{q_{\omega}}{4\pi\,i}\frac{q_{\omega}^{2}}{\Gamma^{2}}\left\{
\int_{0}^{1}{d}{y}\,y\,J_{0}(\rho y)\,\left[\frac{e^{i\,\mid s \mid \sqrt{1-y^{2}}}}{\frac{q_{\omega}}{\Gamma}+\sqrt{1-y^{2}}}\right]\right\}
\nonumber\\
&-&\frac{q_{\omega}}{4\pi\,i}\frac{q_{\omega}^{2}}{\Gamma^{2}}\left\{
i\,
\int_{1}^{\infty}{d}{y}\,y\,J_{0}(\rho y)\,\left[\frac{e^{-\,\mid s \mid \sqrt{y^{2}-1}}}{-i\,\frac{q_{\omega}}{\Gamma}+\sqrt{y^{2}-1}}\right]\right\},
\nonumber\\
\end{eqnarray}
\normalsize
\small
\begin{eqnarray}\label{A3.29}
{G}_{fs}^{yy}(\vec{r}_{\parallel},0;z,0;\omega)& = &\,-\,
\frac{q_{\omega}}{4\pi\,i}\left\{\int_{0}^{1}\frac{{d}{y}\,y\,J_{0}(\rho y)\,e^{i\,\mid s \mid \sqrt{1-y^{2}}}}{\frac{\Gamma}{q_{\omega}}+\sqrt{1-y^{2}}}\,\right\}
\nonumber\\
&+&\,
\frac{q_{\omega}}{4\pi\,i}\left\{
i\,\int_{1}^{\infty}\frac{{d}{y}\,y\,J_{0}(\rho y)\,e^{-\mid s \mid \sqrt{y^{2}-1}}}{-i\,\frac{\Gamma}{q_{\omega}}+\sqrt{y^{2}-1}}\right\},
\nonumber\\
\end{eqnarray}
\begin{eqnarray}\label{A3.30}
{G}_{fs}^{zz}(\vec{r}_{\parallel},0;z,0;\omega)& = &\,-\,
 \frac{q_{\omega}}{4\pi\,i}\frac{q_{\omega}}{\Gamma}\int_{0}^{1}{d}{y}\,y\,J_{0}(\rho y)\,e^{i\,\mid s \mid \sqrt{1-y^{2}}}
\nonumber\\
&\times &
\left\{\frac{y^{2}}{y^{2}+\alpha_{2}\sqrt{1-y^{2}}}\right\}
\nonumber\\
&-&
\frac{q_{\omega}}{4\pi\,i}\frac{q_{\omega}}{\Gamma}\int_{1}^{\infty}{d}{y}\,y\,J_{0}(\rho y)\,e^{-\mid s \mid \sqrt{y^{2}-1}}
\nonumber\\
&\times &
\left\{\frac{y^{2}}{y^{2}+i\,\alpha_{2}\sqrt{y^{2}-1}}\right\}.
\nonumber\\
\end{eqnarray}
\normalsize
where $ \alpha_{0}= \frac{2\,\delta(0)}{q_{\omega}}\sim \frac{2}{d\,q_{\omega}}$, $ \alpha_{1}= \frac{q_{\omega}}{\Gamma}$
and $ \alpha_{2}= \alpha_{1}+\,i\,\alpha_{0}$ and we note that $\delta(z)=0 $ for $ z\neq 0 $.

Furthermore,
\small
\begin{eqnarray}\label{A3.31}
{G}_{fs}^{zx}(\vec{r}_{\parallel},0;z,0;\omega)& = &
 \,\frac{sgn(z)}{4\pi\,i\,}\,\frac{q_{\omega}^{2}}{\Gamma}
 \nonumber\\
&\times &
 \left\{\int_{0}^{1}
\frac{{d}{y}\,y^{2}\,J_{0}(\rho\,y)\,
e^{i\,\mid s \mid\,\sqrt{1-y^{2}}}}{\frac{q_{\omega}}{\Gamma}+\sqrt{1-y^{2}}}\right\}
\nonumber\\
 &-&\,\frac{sgn(z)}{4\pi\,i\,}\,\frac{q_{\omega}^{2}}{\Gamma}
 \nonumber\\
&\times &
 \left\{
\,i\,\int_{1}^{\infty}
\frac{{d}{y}\,y^{2}\,J_{0}(\rho\,y)\,
e^{-\,\mid s \mid\,\sqrt{y^{2}-1}}}{-i\,\frac{q_{\omega}}{\Gamma}+\sqrt{y^{2}-1}}
\right\},
\nonumber\\
\end{eqnarray}
\begin{eqnarray}\label{A3.32}
{G}_{fs}^{xz}(\vec{r}_{\parallel},0;z,0;\omega)& = &
 \,\frac{sgn(z)}{4\pi\,i\,}\frac{q_{\omega}^{2}}{\Gamma}
 \nonumber\\
&\times &
\int_{0}^{1}\frac{{d}{y}\,y^{2}\,J_{0}(\rho\,y)\,
\sqrt{1-y^{2}}\,e^{i\,|s|\sqrt{1-y^{2}}}}{y^{2}+\alpha_{2}\,\sqrt{1-y^{2}}}
\nonumber\\
&+&\,\frac{sgn(z)}{4\pi\,}\frac{q_{\omega}^{2}}{\Gamma}\,
\nonumber\\
&\times &
\,\int_{1}^{\infty}\frac{{d}{y}\,y^{2}\,J_{0}(\rho\,y)\,
\sqrt{y^{2}-1}\,e^{-\,|s|\sqrt{y^{2}-1}}}{y^{2}+i\,\alpha_{2}\,\sqrt{y^{2}-1}}.
\nonumber\\
\end{eqnarray}
\normalsize
Thus, $\widehat{G}_{fs}(\vec{r}_{\parallel},0;z,0;\omega) $ takes the following form in matrix notation (position representation):
\begin{equation}\label{A3.33}
  {\widehat{{G}}_{fs}}({\vec{r}}_{\parallel},0;z,0;\omega) =
  \begin{bmatrix}
  {G}_{fs}^{xx} & 0 & {G}_{fs}^{xz} \\
  0 & {G}_{fs}^{yy} & 0 \\
  {G}_{fs}^{zx} & 0 & {G}_{fs}^{zz}
 \end{bmatrix}.
\end{equation}
Recalling Eq.(\ref{A3.10}) and Eq.(\ref{A3.26})
\begin{equation}\label{A3.34}
   {\widehat{G}}({\vec{r}}_{\parallel},0;z,0;\omega)\,=\,{\widehat{{G}}_{fs}}({\vec{r}}_{\parallel},0;z,0;\omega){\widehat{M}}^{-1},
\end{equation}
we obtain the non-vanishing elements of $\widehat{G}_{fs}(\vec{r}_{\parallel},0;z,0;\omega) $ in a matrix form as
\tiny
\begin{eqnarray}\label{A3.35}
{\widehat{G}}(\vec{r}_{\parallel},0;z,0;\omega)& = &
\nonumber\\
\begin{bmatrix}
  \frac{G_{fs}^{xx}({\vec{r}}_{\parallel},0;z,0;\omega)}
                    {1\,+\,\beta\,G_{fs}^{xx}(0,0;0,0;\omega)} & 0 & \frac{G_{fs}^{xz}({\vec{r}}_{\parallel},0;z,0;\omega)}
                    {1\,+\,\beta\,G_{fs}^{zz}(0,0;0,0;\omega)} \\
  0 &\frac{G_{fs}^{yy}({\vec{r}}_{\parallel},0;z,0;\omega)}
                    {1\,+\,\beta\,G_{fs}^{yy}(0,0;0,0;\omega)} & 0 \\
  \frac{G_{fs}^{zx}({\vec{r}}_{\parallel},0;z,0;\omega)}
                    {1\,+\,\beta\,G_{fs}^{xx}(0,0;0,0;\omega)} & 0 & \frac{G_{fs}^{zz}({\vec{r}}_{\parallel},0;z,0;\omega)}
                    {1\,+\,\beta\,G_{fs}^{zz}(0,0;0,0;\omega)}
 \end{bmatrix}.
 \nonumber\\
\end{eqnarray}
\normalsize
This represents the dyadic Green's function of the system in the presence of the aperture, which is needed to calculate
its electric field response.  It is clear that the plasmonic-polariton dispersion relation is determined by the frequency poles of the matrix elements of the dyad $ {\widehat{G}}({\vec{r}}_{\parallel},0;z,0;\omega)$,
i.e., the zeroes of $ det[{{\widehat{I}}\,+\,\beta\,{\widehat{G}}_{fs}(0,0;0,0;\omega)}]$.

\section{Electrodynamics Wave Transmission of a Perforated Screen}
\begin{figure}[h]
\centering
\includegraphics[width=6cm,height=6cm]{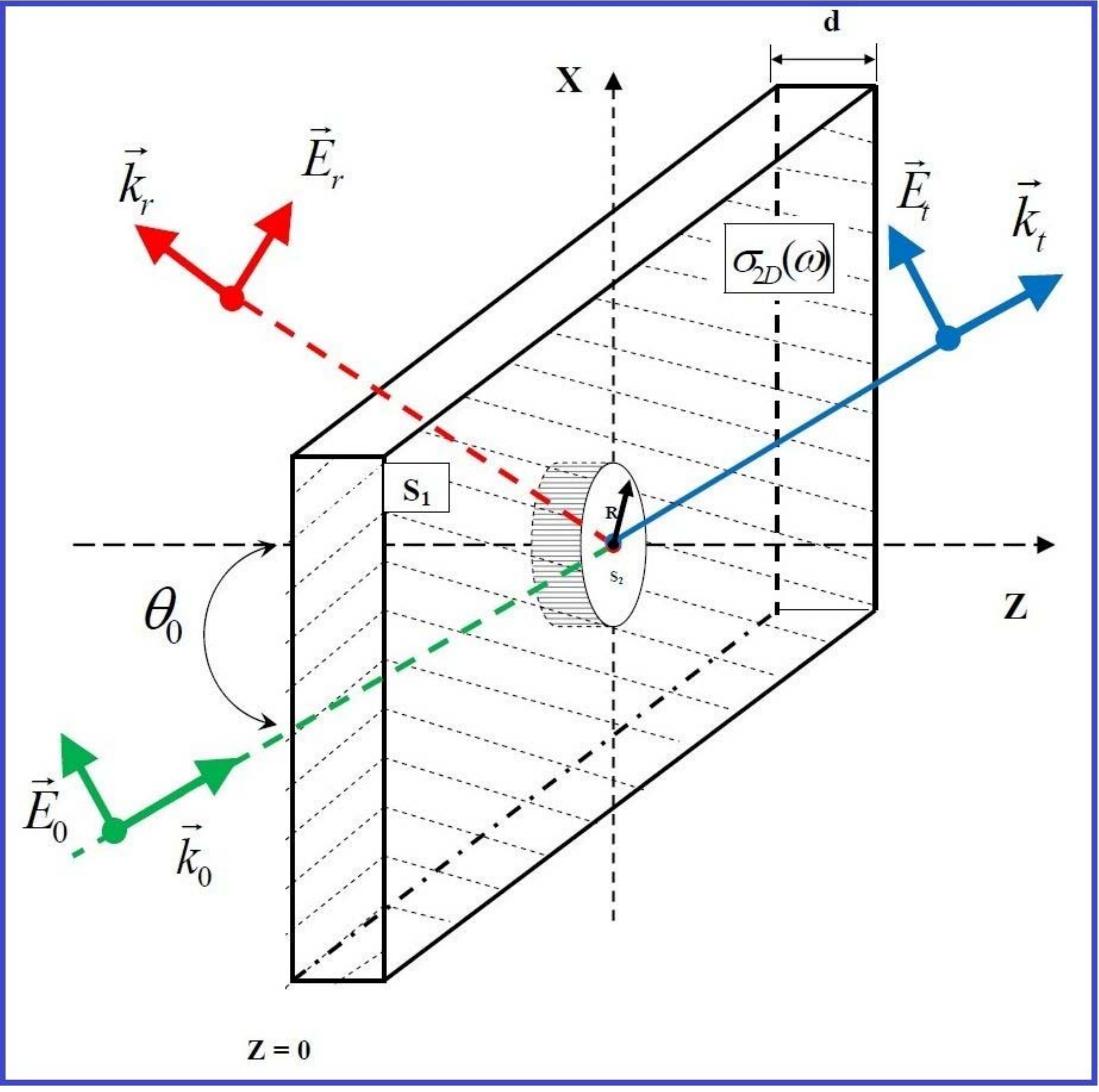}
\caption{Two dimensional plasmonic layer (thickness $ d $, embedded
at $ z=0 $ in a three dimensional bulk medium) with a nano-hole of radius
$ R $ at the origin of the $(x-y)$ plane., shown with incident, reflected and
transmitted wave vectors ($ \vec{k}_{0}, \vec{k}_{r}, \vec{k}_{t} $ ) for waves
($ \vec{E}_{0}(x,y,z;t), \vec{E}_{r}(x,y,z;t),
\newline
\vec{E}_{t}(x,y,z;t)$).}
\label{FIG1411}
\qquad
\end{figure}

    Considering the presence of a nano-hole in a plasmonic screen subject to an external current density source $ \vec{J}_{ext}(\vec{r};t)$,
it is clear that the removal of the material responsible for the introduction of the hole results in the dyadic electromagnetic wave equation
given by
\begin{eqnarray}\label{A4.1}
\left[\widehat{I}\left(\vec{\nabla}^{2}+
\frac{\omega^{2}}{c^{2}}\varepsilon_{b}^{(3D)}\right)
- \vec{\nabla}\vec{\nabla}\right]\vec{E}(\vec{r},\omega)
\nonumber\\
+ \frac{4\pi\,i\omega}{c^{2}}
\int d^{3}\vec{r}^{\,\prime}{\widehat{\sigma}_{fs}}^{(2D)}(\vec{r},\vec{r}^{\,\prime};\omega)
\vec{E}(\vec{r}^{\,\prime};\omega)
\nonumber\\
-
\frac{4\pi\,i\omega}{c^{2}}
\int d^{3}\vec{r}^{\,\prime}{\widehat{\sigma}_{hole}}^{(2D)}(\vec{r},\vec{r}^{\,\prime};\omega)
\vec{E}(\vec{r}^{\,\prime};\omega)
\nonumber\\
=
-\frac{4\pi\,i\omega}{c^{2}}\vec{J}_{ext}(\vec{r};\omega),
\end{eqnarray}
where $ \widehat{\sigma}_{fs}^{(2D)}$ is the conductivity tensor of the 2D layer and
$ \widehat{\sigma}_{hole}^{(2D)} $ is the conductivity tensor of the nano-hole.
In this context, the dyadic Green's function $ \widehat{G} $ for the perforated screen is given by the integro-differential equation
\begin{eqnarray}\label{A4.2}
\left[\widehat{I}\left(\vec{\nabla}^{2}+
\frac{\omega^{2}}{c^{2}}\varepsilon_{b}^{(3D)}\right)
- \vec{\nabla}\vec{\nabla}\right]\widehat{G}(\vec{r},\vec{r}^{\,\prime},\omega)
\nonumber\\
+\frac{4\pi\,i\omega}{c^{2}}
\int d^{3}\vec{r}^{'}{\widehat{\sigma}_{fs}}^{(2D)}(\vec{r},\vec{r}^{\,\prime\,\prime};\omega)\,
\widehat{G}(\vec{r}^{\,\prime\,\prime},\vec{r}^{\,\prime},\omega)
\nonumber\\
-
\frac{4\pi\,i\omega}{c^{2}}
\int d^{3}\vec{r}^{\,\prime}{\widehat{\sigma}_{hole}}^{(2D)}(\vec{r},\vec{r}^{\,\prime\,\prime};\omega)
\widehat{G}(\vec{r}^{\,\prime\,\prime},\vec{r}^{\,\prime},\omega)
\nonumber\\
=\,-\,\delta^{(3)}(\vec{r}-\vec{r}^{\,\prime}),
\end{eqnarray}
which can be rewritten in positional matrix notation as
\begin{eqnarray}\label{A4.3}
\widehat{G}= \widehat{G}_{3D}
& + & \frac{4\pi\,i\omega}{c^{2}}\,\widehat{G}_{3D}\left[
{\widehat{\sigma}_{fs}}^{(2D)}\,-\,
{\widehat{\sigma}_{hole}}^{(2D)}\right]
\widehat{G},
\end{eqnarray}
and is equivalent to Eq.(\ref{A3.2}).
Bearing in mind that Green's functions are defined to relate fields to currents, the very distant current source $ \vec{J}_{ext}(\vec{r};\omega)$
of an incident field $ \vec{E}_{0}$ is related to it by
\begin{equation}\label{A4.4}
\vec{E}_{0}= \,+\,\frac{4\pi\,i\,\omega}{c^{2}} \widehat{G}_{3D}\vec{J}_{ext},
\end{equation}
for linear media. Therefore, the total field  $ \vec{E}$ induced by an incident field $ \vec{E}_{0}$ is given by [1],[2]
\begin{eqnarray}\label{A4.5}
\vec{E}&=& \,+\,\frac{4\pi\,i\,\omega}{c^{2}} \widehat{G}\,\vec{J}_{ext}
\nonumber\\
&=&
\widehat{G}\,{\widehat{G}_{3D}}^{-1}\,\vec{E}_{0}
\end{eqnarray}
and following the procedures of the references cited, we obtain
\begin{equation}\label{A4.6}
\vec{E} = \vec{E}_{0} \,+\, \frac{4\pi\,i\,\omega}{c^{2}}\,\widehat{G}
\,\left[
{\widehat{\sigma}_{fs}}^{(2D)}\,-\,
{\widehat{\sigma}_{hole}}^{(2D)}\right]\,\vec{E}_{0},
\end{equation}
which includes both the transmitted and reflected fields.

Alternatively, Eq.(\ref{A4.6}) may be written as
\begin{eqnarray}\label{A4.7}
\vec{E}(\vec{r}_{\parallel};z;\omega) =
\vec{E}_{0}(\vec{r}_{\parallel};z;\omega)+\vec{E}_{1}(\vec{r}_{\parallel},z;\omega)+\vec{E}_{2}(\vec{r}_{\parallel},z;\omega),
\nonumber\\
\end{eqnarray}
where the electric field contributions $\vec{E}_{1}$ and $\vec{E}_{2}$ are define by
\begin{eqnarray}\label{A4.8}
\vec{E}_{1}(\vec{r}_{\parallel},z;\omega)& =&
\frac{4\pi\,i\,\omega}{c^{2}}\int{d^{2}\vec{r}_{\parallel}^{'}}
\int{d^{2}\vec{r}_{\parallel}^{''}} \int{dz^{'}}
\nonumber\\
&\times&
\int{dz^{''}}
\widehat{G}(\vec{r}_{\parallel},\vec{r}_{\parallel}^{'};z,z^{'};\omega)
\nonumber\\
&\times&
{\widehat{\sigma}_{fs}}^{(2D)}
(\vec{r}_{\parallel}^{'},\vec{r}_{\parallel}^{''};z^{'},z^{''};\omega)\,
\vec{E}_{0}(\vec{r}_{\parallel}^{''};z^{''};\omega)
\nonumber\\
\end{eqnarray}
and
\begin{eqnarray}\label{A4.9}
\vec{E}_{2}(\vec{r}_{\parallel},z;\omega) &=&
\frac{4\pi\,i\,\omega}{c^{2}}\int{d^{2}\vec{r}_{\parallel}^{'}}
\int{d^{2}\vec{r}_{\parallel}^{''}}
\nonumber\\
&\times&
\int{dz^{'}} \int{dz^{''}}
\widehat{G}(\vec{r}_{\parallel},\vec{r}_{\parallel}^{'};z,z^{'};\omega)
\nonumber\\
&\times&
\,
{\widehat{\sigma}_{hole}}^{(2D)}
(\vec{r}_{\parallel}^{'},\vec{r}_{\parallel}^{''};z^{'},z^{''};\omega)
\vec{E}_{0}(\vec{r}_{\parallel}^{''};z^{''};\omega).
\nonumber\\
\end{eqnarray}
Employing Eq.(\ref{A3.5}) and executing all the integrations, $ \vec{E}_{2}(\vec{r}_{\parallel},z;\omega)$ reduces to
\begin{eqnarray}\label{A4.10}
\vec{E}_{2}(\vec{r}_{\parallel},z;\omega) &=&
\beta\,\widehat{G}(\vec{r}_{\parallel},0;z,0;\omega)
\,\vec{E}_{0}(0,0;\omega).
\end{eqnarray}

Furthermore, Eq.(\ref{A2.8}) in position space, given by
\small
\begin{equation}\label{A4.11}
   \widehat{\sigma}_{f_{s}}^{2D}(\vec{r}_{\parallel}^{'},\vec{r}_{\parallel}^{''};z^{'},z^{''};\omega)=
\widehat{I}\, {\sigma}_{fs}^{(2D)}(\omega)\delta^{(2D)}
(\vec{r}_{\parallel}^{'}-\vec{r}_{\parallel}^{''})\,\delta(z^{'})\,\delta(z^{''})
\end{equation}
\normalsize
leads to
\small
\begin{eqnarray}\label{A4.12}
\vec{E}_{1}(\vec{r}_{\parallel},z;\omega)& =&
\gamma\,\int{d^{2}\vec{r}_{\parallel}^{'}}
\widehat{G}(\vec{r}_{\parallel},\vec{r}_{\parallel}^{'};z,0;\omega)
\,\vec{E}_{0}(\vec{r}_{\parallel}^{'};0;\omega).
\nonumber\\
\end{eqnarray}
\normalsize
Considering $ z^{'}=0$ in Eq.(\ref{A3.9}) and substituting it into Eq.(\ref{A4.12}), we find
\small
\begin{eqnarray}\label{A4.13}
E_{1}(\vec{r}_{\parallel},z;\omega)& =&
\gamma\,\int{d^{2}\vec{r}_{\parallel}^{'}}
\widehat{G}_{fs}(\vec{r}_{\parallel},\vec{r}_{\parallel}^{'};z,0;\omega)
\,\vec{E}_{0}(\vec{r}_{\parallel}^{'};0;\omega)
\nonumber\\
&-&
\gamma\,
{\beta \widehat{G}_{fs}(\vec{r}_{\parallel},0;z,0;\omega)}
\nonumber\\
&\times&
\left[{\widehat{I}\,+\,\beta
\widehat{G}_{fs}(0,0;0,0;\omega)}\right]^{-1}
\nonumber\\
&\times&
\int{d^{2}\vec{r}_{\parallel}^{'}}
\widehat{G}_{fs}(0,\vec{r}_{\parallel}^{'};0,0;\omega)
\,\vec{E}_{0}(\vec{r}_{\parallel}^{'};0;\omega).
\end{eqnarray}
\normalsize
Thus, the electric field produced by the incident field $ \vec{E}_{0}(\vec{r}_{\parallel};z;\omega)$ in the presence of the nano-hole is described by
\begin{eqnarray}\label{A4.14}
\vec{E}(\vec{r}_{\parallel};z;\omega) &=&
\vec{E}_{0}(\vec{r}_{\parallel};z;\omega)
\nonumber\\
&\,+\,&
\vec{E}_{1}(\vec{r}_{\parallel},z;\omega)
\nonumber\\
&\,-\,& \beta\,
\widehat{G}(\vec{r}_{\parallel},0 ;z,0;\omega)\vec{E}_{0}(0;0;\omega),
\nonumber\\
\end{eqnarray}
where $ \vec{E}_{1}(\vec{r}_{\parallel},z;\omega)$ , however, needs to be determined in explicit form.

Considering an incident external transverse plane electromagnetic wave  $ \vec{E}_{0}(\vec{r};t)$ impinging on the perforated
2D plasmonic layer at an angle of incidentce $ \theta_{0}$ with frequency $ \omega_{0} $, we have
\begin{eqnarray}\label{A4.15}
\vec{E}_{0}(\vec{r};t) &=& \vec{E}_{0} e^{i[\vec{k}_{0}.{\vec{r}}\,-\,{\omega_{0} t}]}
\nonumber\\
&=&
 \vec{E}_{0} e^{i[\vec{k}_{0_{\parallel}}.{\vec{r}_{\parallel}}\,+\, {k_{{0}_{z}}} z\, -\, {\omega_{0} t}]}
\end{eqnarray}
where $ \vec{E}_{0} $ is a vector amplitude perpendicular to the $\vec{k}_{0}$ wave vector(having otherwise unspecified polarization).
In frequency representation,
\begin{equation}\label{A4.16}
\vec{E}_{0}(\vec{r}_{_{\parallel}};z;\omega) = \vec{\widetilde{E}}_{0} e^{i[\vec{k}_{0_{\parallel}}.{\vec{r}_{\parallel}}\,+\, {k_{{0}_{z}}} z]}
\end{equation}
where $ \vec{\widetilde{E}}_{0}=2\pi\delta(\omega-\omega_{0})\vec{E}_{0} $ and $ k_{0_{z}}= \sqrt{q_{\omega_{0}}^{2}-k_{0_{\parallel}}^{2}}$.
Here, $ q_{\omega_{0}} = \frac{\omega_{0}}{c} \sqrt{\varepsilon_{b}^{(3D)}}$  is the incident wavenumber and, $ k_{0_{\parallel}}$ and $ k_{0_{z}}$
are the parallel and perpendicular wave vector components of the incident wavevector, respectively. Employing the plane wave of Eq.(\ref{A4.16}) in Eq.(\ref{A4.13}),
we evaluate the spatial Fourier transform
\begin{eqnarray}\label{A4.17}
\vec{E}_{1}(\vec{r}_{\parallel},z;\omega)& =&
\gamma\,
\widehat{{G}}_{fs}(\vec{k}_{0_{\parallel}};z,0;\omega)\,\vec{\widetilde{E}}_{0}\,
e^{i\,\vec{k}_{0_{\parallel}}\,\cdot\,\vec{r}_{\parallel}}\,
\nonumber\\
&-&
\gamma\,
{\beta \widehat{G}_{fs}(\vec{r}_{\parallel},0;z,0;\omega)}
\nonumber\\
&\times&
\left[{\widehat{I}+\beta
\widehat{G}_{fs}(0,0;0,0;\omega)}\right]^{-1}
\nonumber\\
&\times&
\,\widehat{{{G}}}_{fs}(\vec{k}_{0_{\parallel}};0,0;\omega)\,\vec{\widetilde{E}}_{0}\,,
\end{eqnarray}
which yield the electric field of Eq.(\ref{A4.14}) (in time domain) as
\small
\begin{eqnarray}\label{A4.18}
\vec{E}(\vec{r}_{\parallel};z;t)&=&
\vec{E}_{0}(\vec{r}_{\parallel};z;t)
\nonumber\\
&\,+\,&
\gamma_{0}\,
\widehat{{G}}_{fs}(\vec{k}_{0_{\parallel}};z,0;\omega_{0})\,\vec{E}_{0}\,
e^{i\,\left[\,\vec{k}_{0_{\parallel}}\,\cdot\,\vec{r}_{\parallel}\,-\,\omega_{0}\,t\right]}\,
\nonumber\\
&-&
\,\beta_{0}\,\widehat{G}(\vec{r}_{\parallel},0;z,0;\omega_{0})
\nonumber\\
&\times&
\left[{\widehat{I}+\gamma_{0}
\widehat{\overline{\overline{G}}}_{fs}(\vec{k}_{0_{\parallel}};0,0;\omega_{0})}\right]
\,\vec{E}_{0}\,\,e^{-i\,\omega_{0}\,t}.
\end{eqnarray}
\normalsize
Here, $ \beta_{0} = \gamma_{0}\, {A}^{2} $ and with Eq.(\ref{A3.4}) we have $ \sigma_{fs}^{(2D)}(\omega_{0})=\frac{i\,d\,\omega^{2}_{P_{3D}}}{4\pi\,\omega_{0}}$
where the local bulk plasma polarizability $ \sim \,-\,\frac{\omega_{P3D}^{2}}{\omega_{0}^{2}}$ is used and $ \omega_{_{P3D}}= \sqrt{\frac{4\,\pi\,e^{2}\,\rho_{_3D}}{m^{\star}}}$
is the 3D bulk plasma frequency.  With this, Eq.(\ref{A2.11}) yields
\begin{equation}\label{A4.19}
\gamma_{0}=\,-\,\frac{d\,\omega_{P_{3D}}^{2}}{c^{2}}\,\,\,\,\, \textrm{and} \,\,\,\,\Gamma_{0}=\,-\,\frac{i\,\gamma_{0}}{2}.
\end{equation}

Recognizing that the first two terms on the right of Eq.(\ref{A4.18}) is just $ \vec{E}_{fs}(\vec{r}_{\parallel};z;t)$, we define the dyadic function $ \widehat{T}^{0}$ as
\begin{equation}\label{A4.20}
 \widehat{T}^{0}=\widehat{G}_{0}(\vec{r}_{\parallel},0;z,0;\omega_{0})
\,\left[\widehat{I}\,+\,\gamma_{0}\,\widehat{{G}}_{fs}(\vec{k}_{0_{\parallel}};0,0;\omega_{0})\right],
\end{equation}
so that
\begin{equation}\label{A4.21}
\vec{E}(\vec{r}_{\parallel};z;t) = \vec{E}_{fs}(\vec{r}_{\parallel};z;t)\, - \,
\beta_{0} \, \widehat{T}^{0}
\vec{E}_{0}\,e^{-i\,\omega_{0}t}.
\end{equation}
Since the incident wavervector $ \vec{k_{0}}$ is in the $(x-z)$-plane of incidence ($ k_{0y}=0$) and the dyad
$ \widehat{{G}}_{fs}(\vec{k}_{0_{\parallel}};0,0;\omega_{0})$ was found to be diagonal (Eq.(\ref{A3.12})), then Eq.(\ref{A4.20}) for $ \widehat{T}^{0}$
takes a matrix form fully described in Appendix A3.

Substituting Eq.(\ref{D.2}) into Eq.(\ref{A4.21}) using the matrix elements of Appendix A3 leads to $ \vec{E}(x,y,z;t)$ in matrix form
\begin{eqnarray}\label{A4.22}
\begin{bmatrix}
  E_{x}(x,y,z;t)\\
  E_{y}(x,y,z;t)\\
  E_{z}(x,y,z;t)\\
\end{bmatrix}
&=&
\begin{bmatrix}
  E_{fs_{x}}(x,y,z;t)\\
  E_{fs_{y}}(x,y,z;t)\\
  E_{fs_{z}}(x,y,z;t)\\
\end{bmatrix}
\nonumber\\
&\,-\,&\beta_{0}
\begin{bmatrix}
  {T}_{xx}^{0} & 0 & {T}_{xz}^{0} \\
  0 & {T}_{yy}^{0} & 0 \\
  {T}_{zx}^{0} & 0 & {T}_{zz}^{0}
 \end{bmatrix}
 \nonumber\\
&\times&
\begin{bmatrix}
  E_{0_{x}}\,e^{-i\omega_{0} t}\\
  E_{0_{y}}\,e^{-i\omega_{0} t}\\
  E_{0_{z}}\,e^{-i\omega_{0} t}\\
\end{bmatrix},
\nonumber\\
\end{eqnarray}
or
\begin{eqnarray}\label{A4.23}
             E_{x}(x,y,z;t) &=&  E_{fs_{x}}(x,y,z;t)
             \nonumber\\
             &\,-\,&\beta_{0}\left[ {T}_{xx}^{0}\,E_{0_{x}}+{T}_{xz}^{0}\,E_{0_{z}}\right]\,e^{-i\,\omega_{0} t}  ;
\end{eqnarray}
\begin{eqnarray}\label{A4.23I}
             E_{y}(x,y,z;t)& =&  E_{fs_{y}}(x,y,z;t)
             \nonumber\\
             &\,-\,&\beta_{0}\left[ {T}_{yy}^{0}\,E_{0_{y}}\right] \,e^{-i\,\omega_{0} t} ;
\end{eqnarray}
\begin{eqnarray}\label{A4.23II}
E_{z}(x,y,z;t) &= & E_{fs_{z}}(x,y,z;t)
\nonumber\\
 &\,-\,&\beta_{0}\left[ {T}_{zx}^{0}\,E_{0_{x}}+{T}_{zz}^{0}\,E_{0_{z}}\right]\,e^{-i\,\omega_{0} t}
\end{eqnarray}

upon which our computations are based.

\section{Numerical Analysis of Electromagnetic Wave Propagation in the Vicinity of a Perforated 2D Plasmonic Layer for Normal Incidence ($ \theta_{0}=0$)}

The subwavelength restriction imposed on a nano-hole of dimension $ R < \lambda_{0} $ (incident wavelength) requires that
$ q_{\omega_{0}}R < 1 $; in this matter, the incident wave frequency $ \omega_{0}$ must fall below the cutoff frequency $ \widetilde{\omega}_{0}$
given by
\begin{equation}\label{A5.1}
    \omega_{0} < \widetilde{\omega}_{0}\equiv \frac{c}{R\,\sqrt{\varepsilon_{b}^{(3D)}}},\, \textrm{alternatively}\,\, f_{0}=\frac{\omega_{0}}{2\,\pi}<\widetilde{f}_{0}\,=\, \frac{\widetilde{\omega}_{0}}{2\,\pi}.
\end{equation}

Employing Mathematica to evaluate the matrix elements of the dyadics $ \widehat{G}_{fs}(\vec{r}_{\parallel},0;z,0;\omega_{0})$ and $\widehat{T}^{0}$ (of section 4),
we carry out a numerical determination of the electric field generated by $ \vec{E}_{0}$ for normal incidence $(\theta_{0}=0)$.  The transmitted ($ z > 0 $) and reflected ($ z < 0 $)
field distributions in space are presented graphically in terms of the square modulii of the fields.

For normal incidence the transmitted field $ (z > 0)$ distributions are calculated for $ R =5\,nm $, $ d = 10\,nm $ and $ {f}_{0}= 300\,THz $
in terms of parallel $(E_{\parallel})$ and normal $(E_{z})$ components for near-field, middle-field and far-field diffraction zones for $ z=50\,R $, $ z=300\,R $ and $ z=1000\,R $,
respectively.  The figures exhibit $\mid E_{\parallel}(x,y,z;t)/E_{0}\mid^{2}$ and $\mid E_{z}(x,y,z;t)/E_{0}\mid^{2}$ for a perforated $ GaAs $ screen with effective mass $
m^{\ast}=0.067\,m_{0} $ ($ m_{0} $ is the free-electron mass) and density $ n_{3D}=4 \times 10^{21}/cm^{3}$ ($ \varepsilon_{b}^{(3D)}=1 $ is the dielectric constant of the host
medium).  This is in fact clearly illustrated with $ y \equiv 0 $ below.
\newpage
\subsection{Normal Incidence}
\subparagraph{For the Near-Field zone, we take $ z = 50\,R $}.
\begin{figure}[h]
\centering
(a)\includegraphics[width=8.0cm,height=6.0cm]{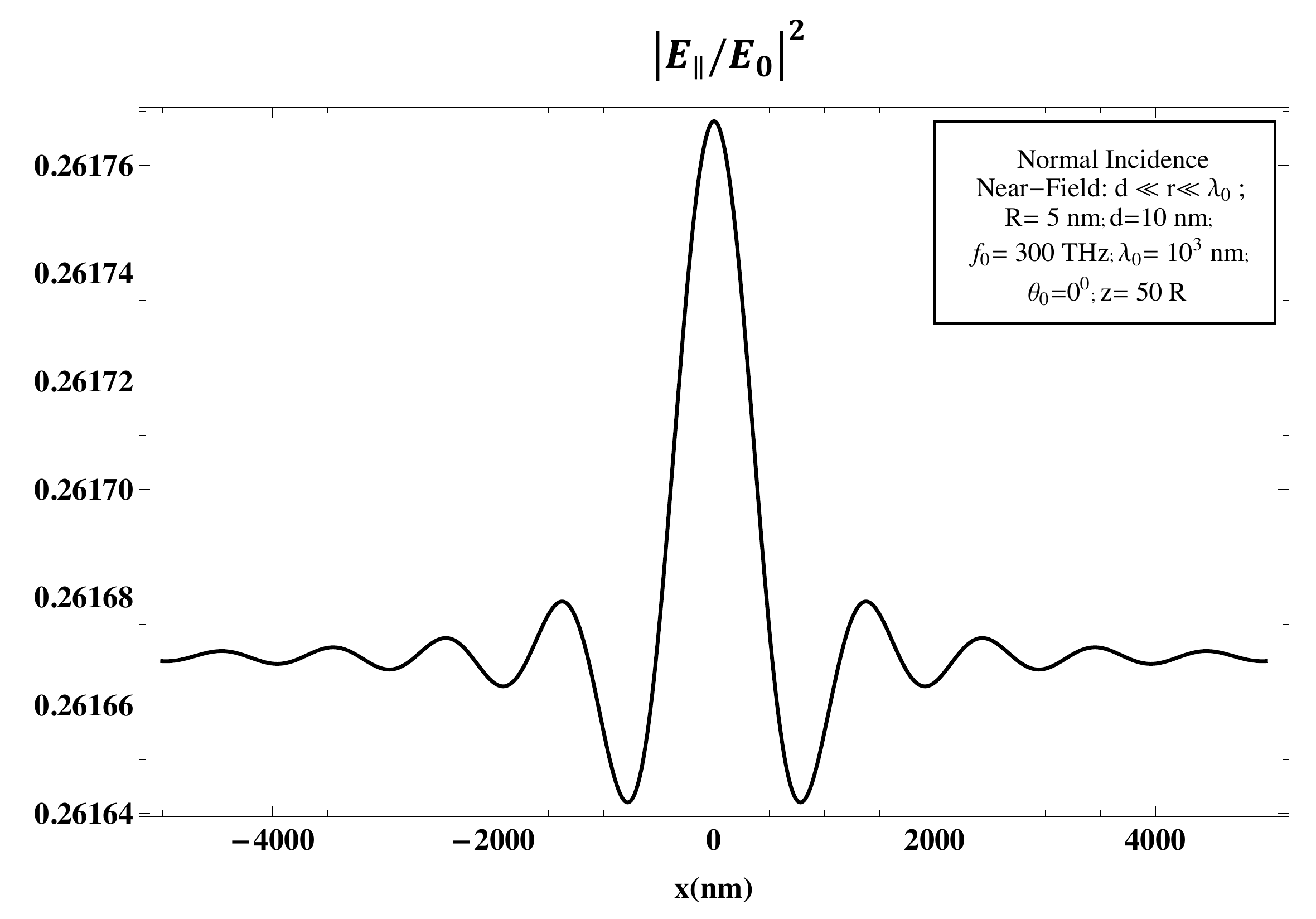}
(b)\includegraphics[width=8.0cm,height=6.0cm]{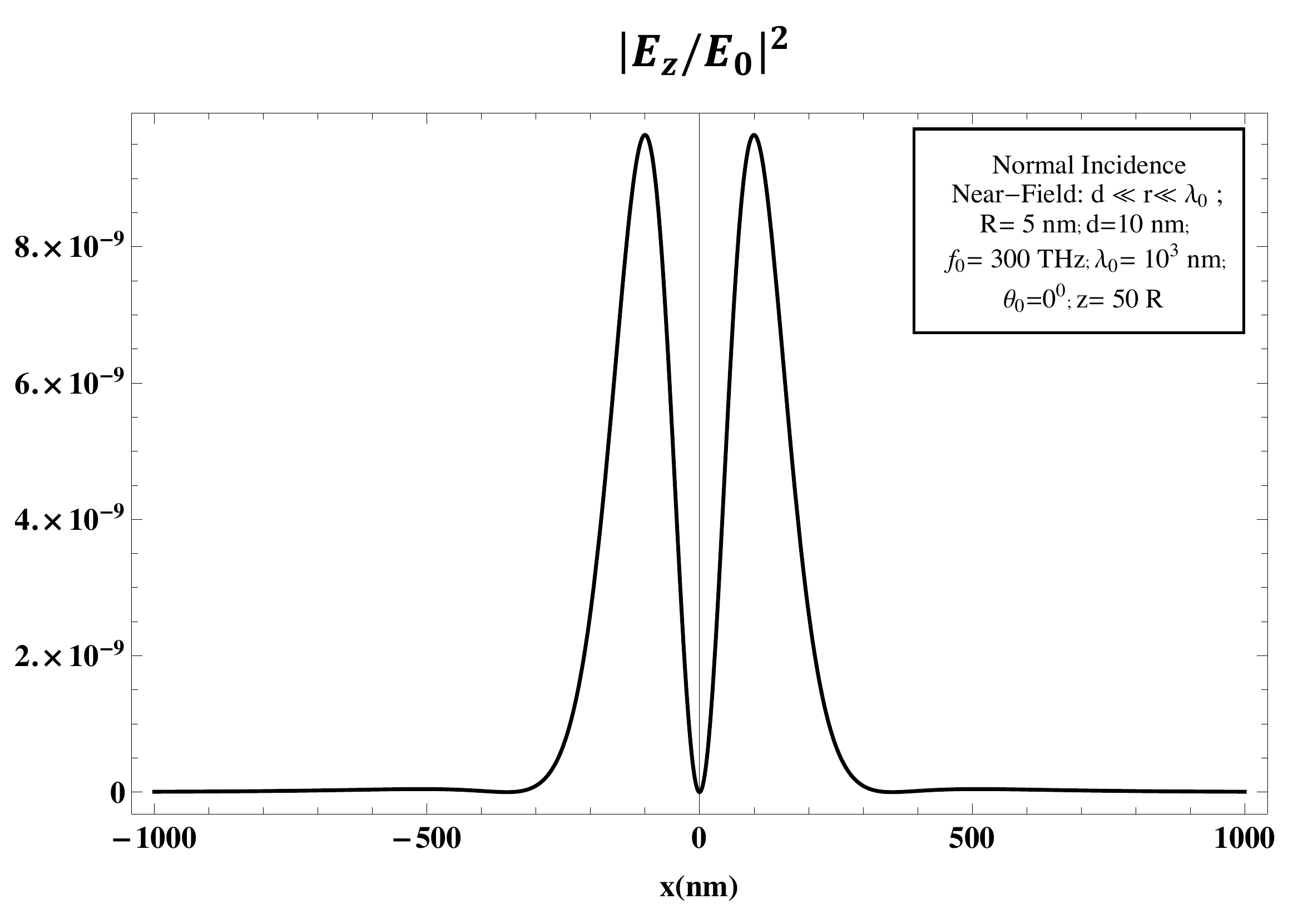}
(c)\includegraphics[width=8.0cm,height=6.0cm]{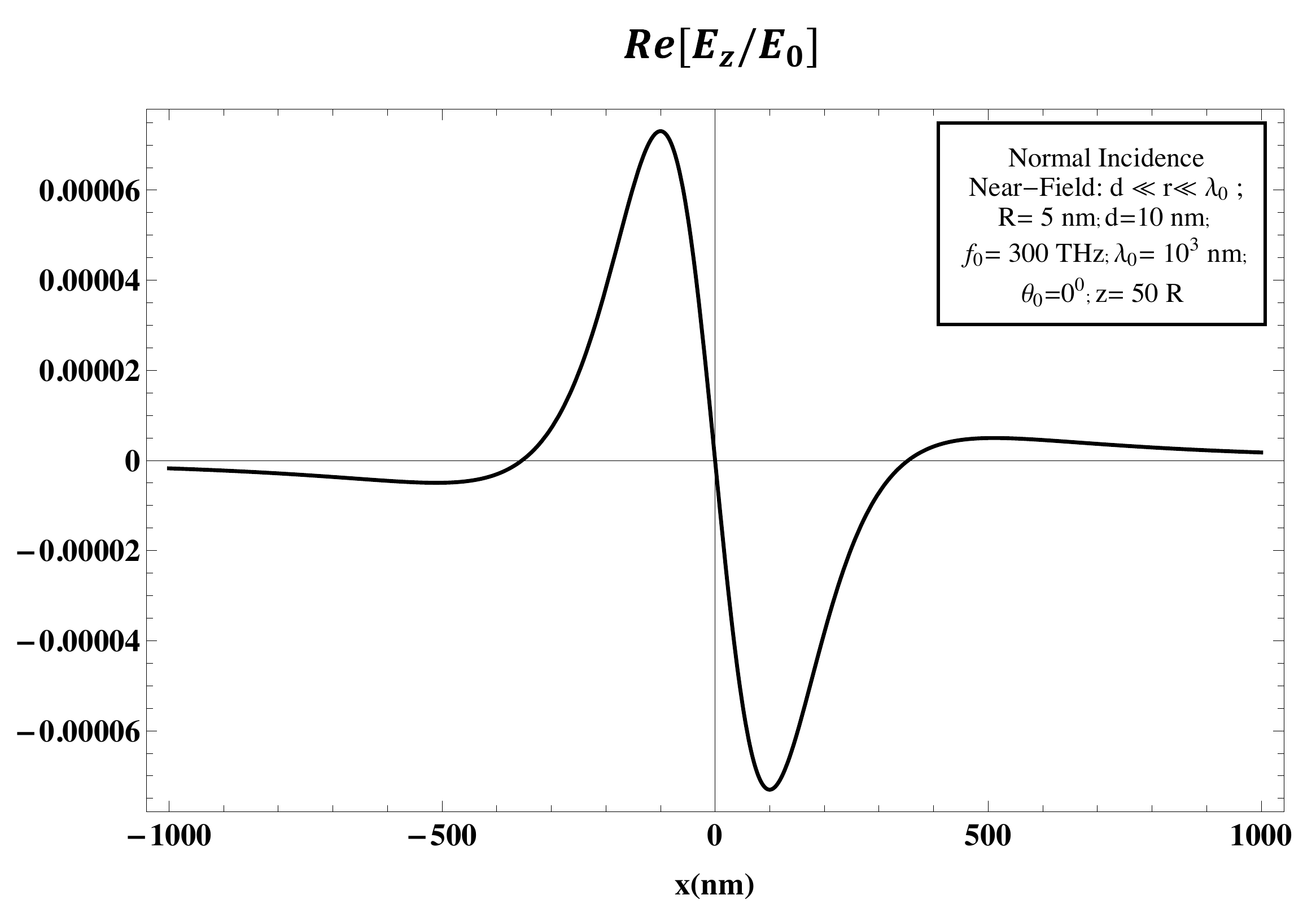}
\caption{$\mid {E}_{\parallel}(x,y,z;t)/{E}_{0}\mid^{2}$(a), $\mid E_{z}(x,y,z;t)/E_{0}\mid^{2}$(b) and $ Re[E_{z}(x,y,z;t)/E_{0}]$(c) produced by a perforated
2D plasmonic layer of GaAs as a function of lateral distance $ r_{_{\parallel}}= x\,(y=0)$ from the aperture.}
\label{FIGNFT0E}
\end{figure}
\newpage
\subparagraph{For the Middle-Field zone: $ z = 300\,R $}.
\begin{figure}[h]
\centering
(a)\includegraphics[width=8.0cm,height=6.0cm]{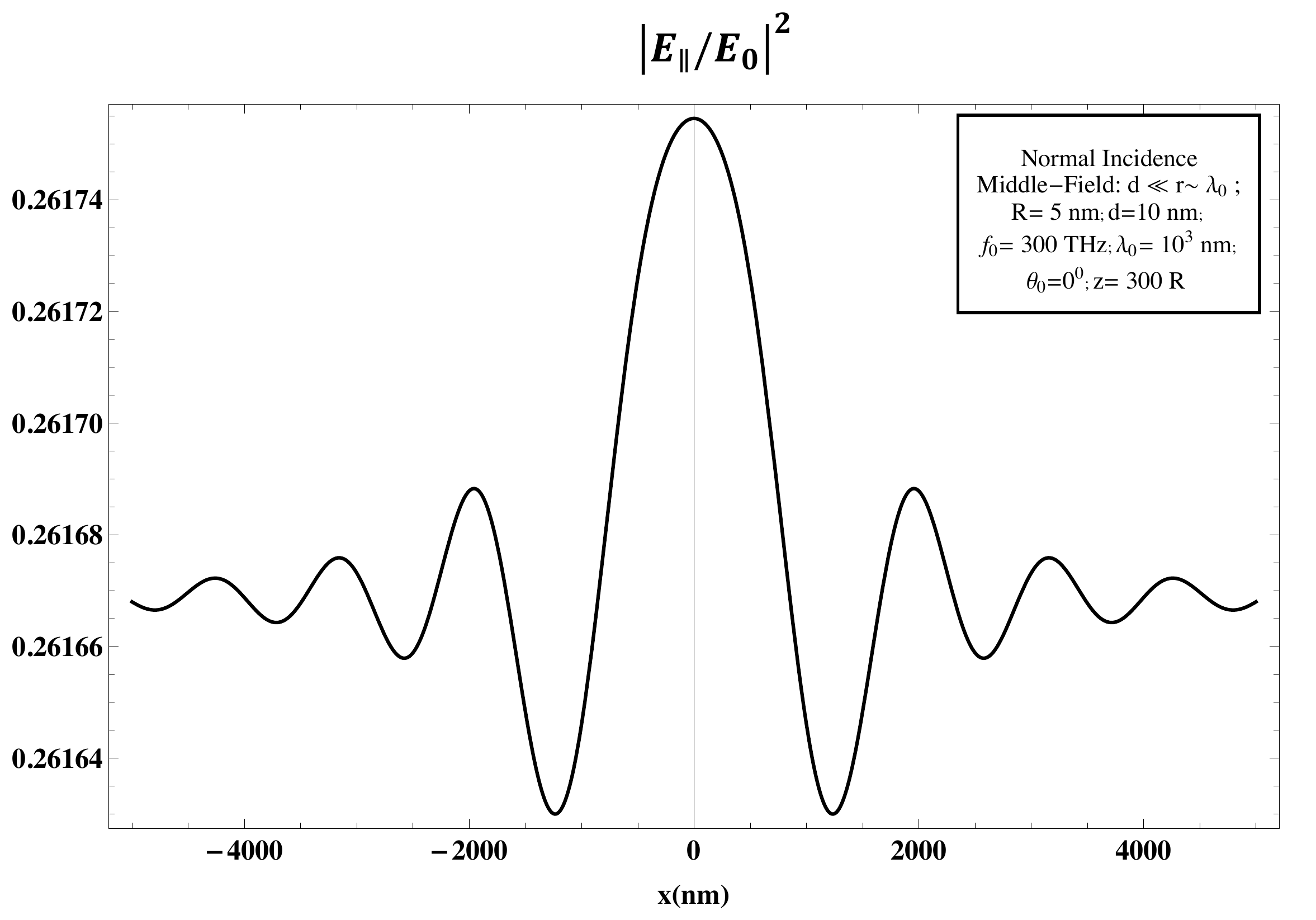}
(b)\includegraphics[width=8.0cm,height=6.0cm]{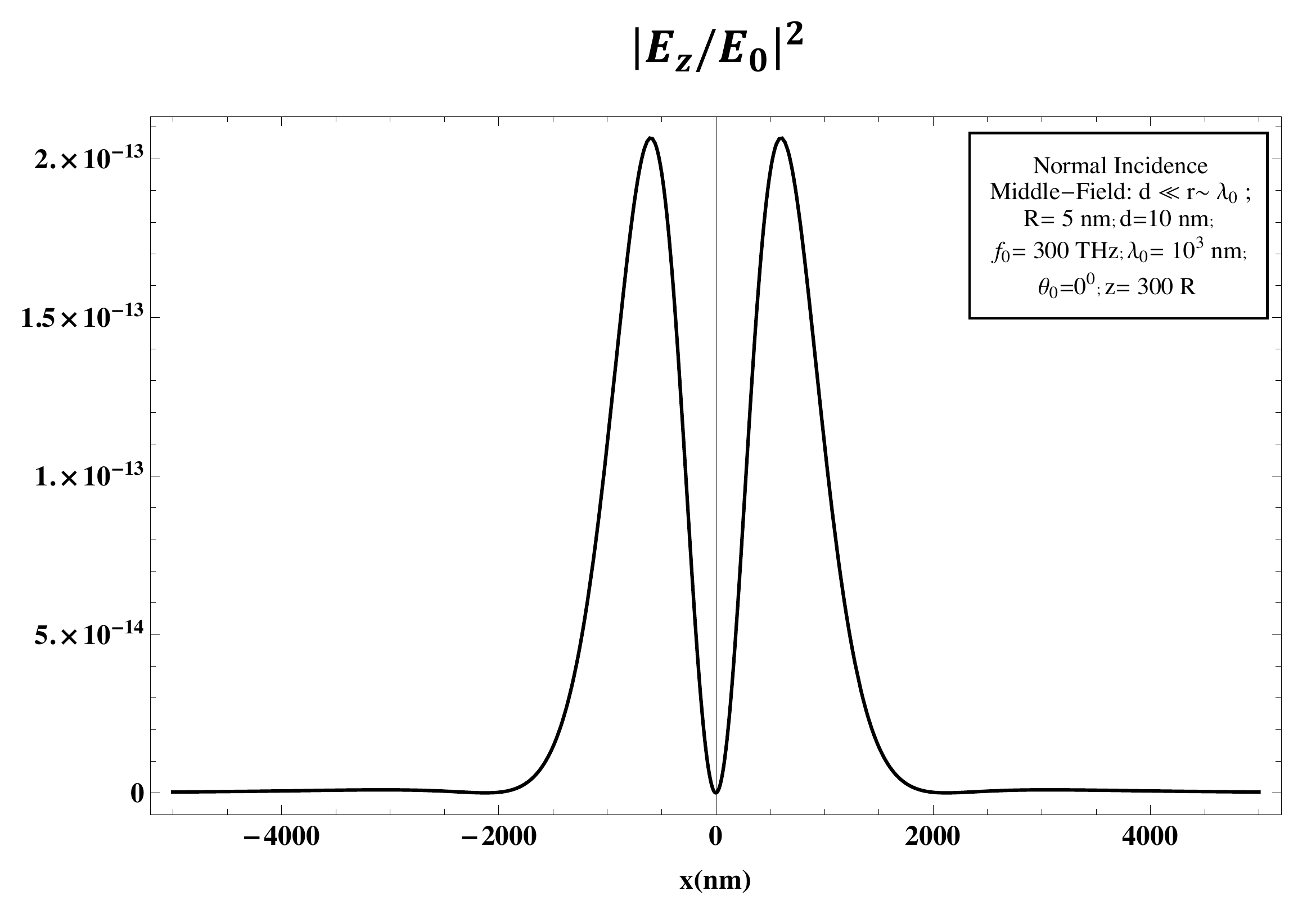}
(c)\includegraphics[width=8.0cm,height=6.0cm]{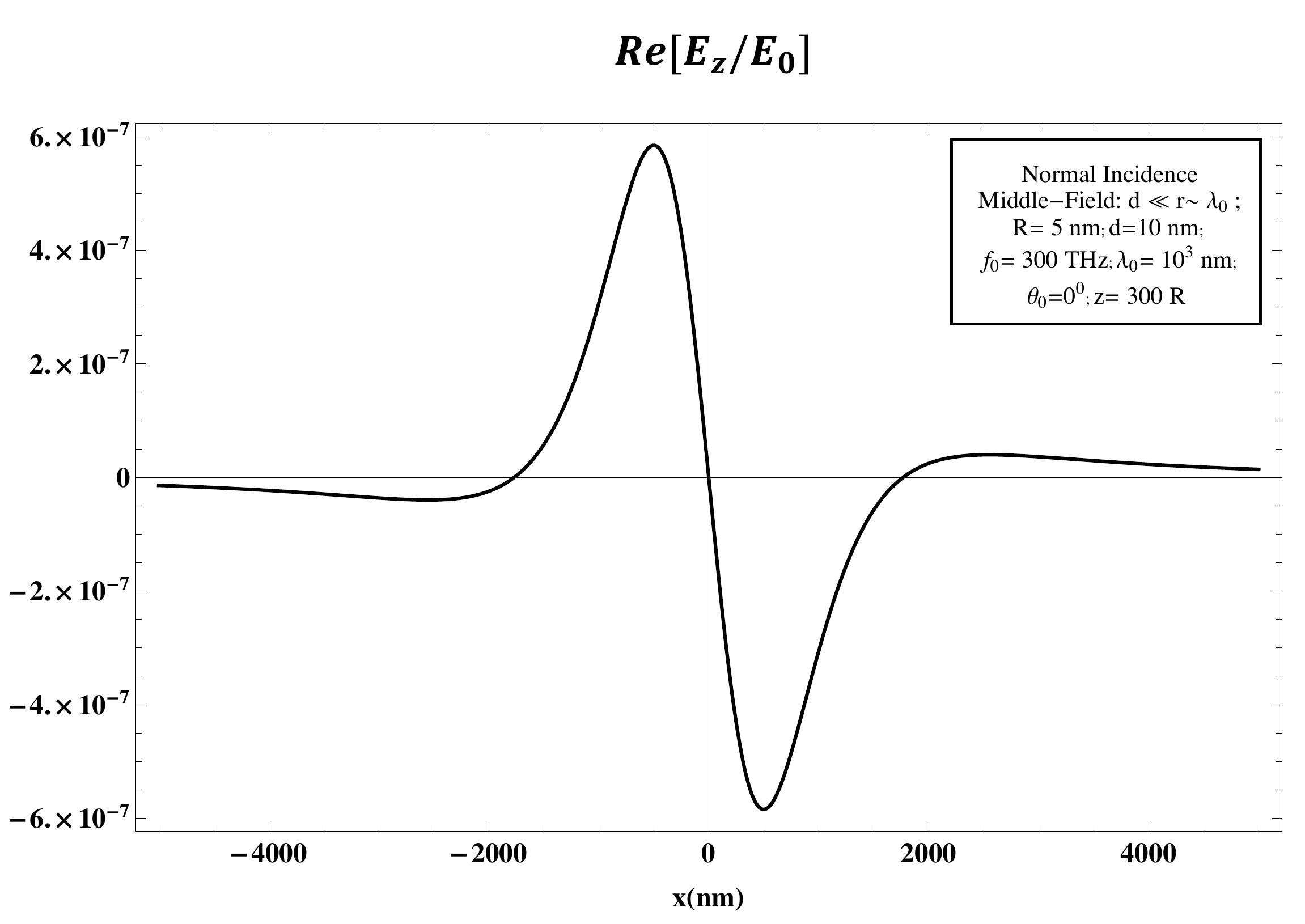}
\caption{ $\mid {E}_{\parallel}(x,y,z;t)/{E}_{0}\mid^{2}$(a), $\mid E_{z}(x,y,z;t)/E_{0}\mid^{2}$(b) and $ Re[E_{x}(x,y,z;t)/E_{0}]$(c) produced by a perforated
2D plasmonic layer of GaAs as a function of lateral distance $ r_{_{\parallel}}= x\,(y=0)$ from the aperture.}
\label{FIGMFT0E}
\end{figure}
\newpage
\subparagraph{For the Far-Field zone: $ z = 1000\,R $}.
\begin{figure}[h]
\centering
(a)\includegraphics[width=8.0cm,height=6.0cm]{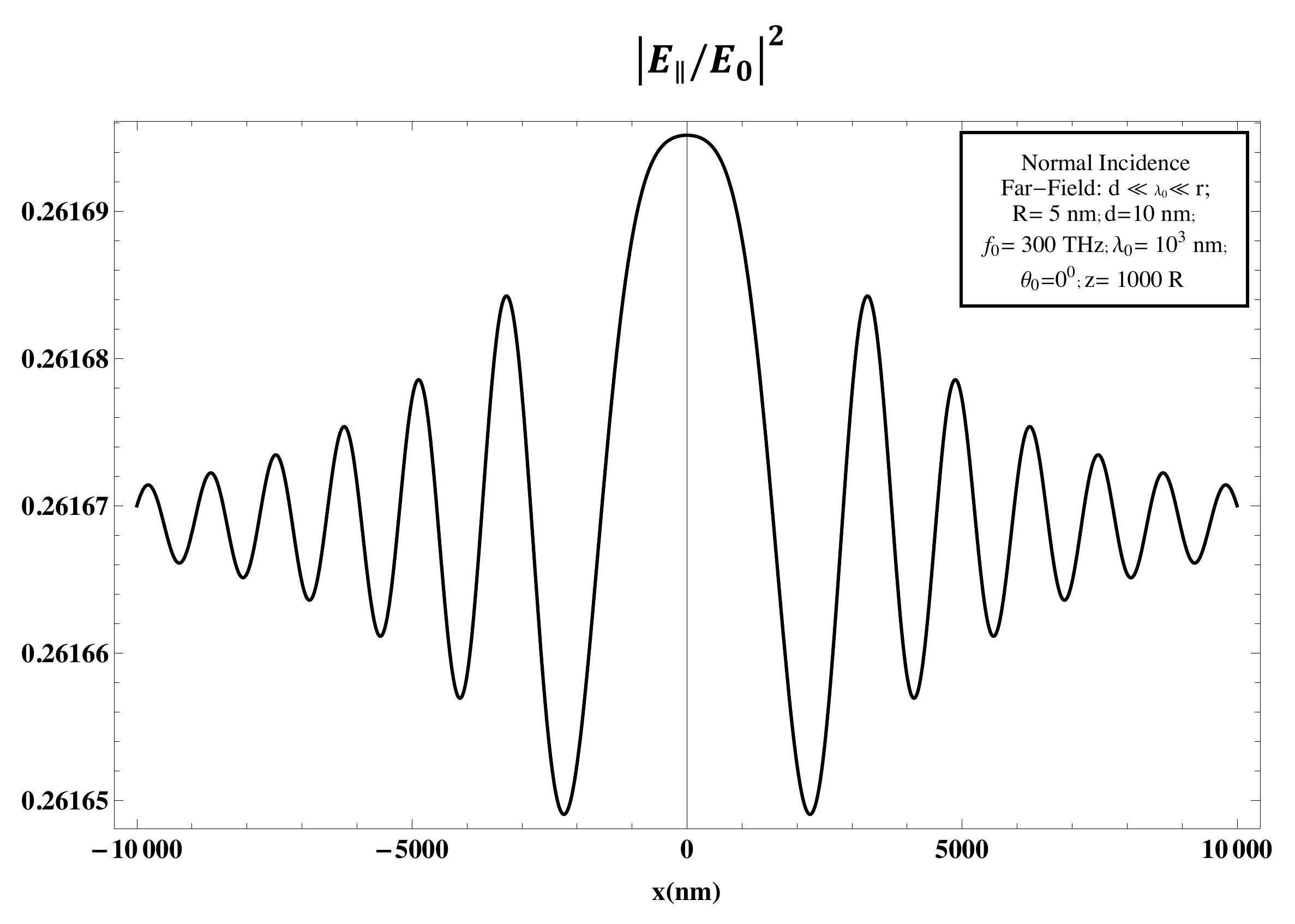}
(b)\includegraphics[width=8.0cm,height=6.0cm]{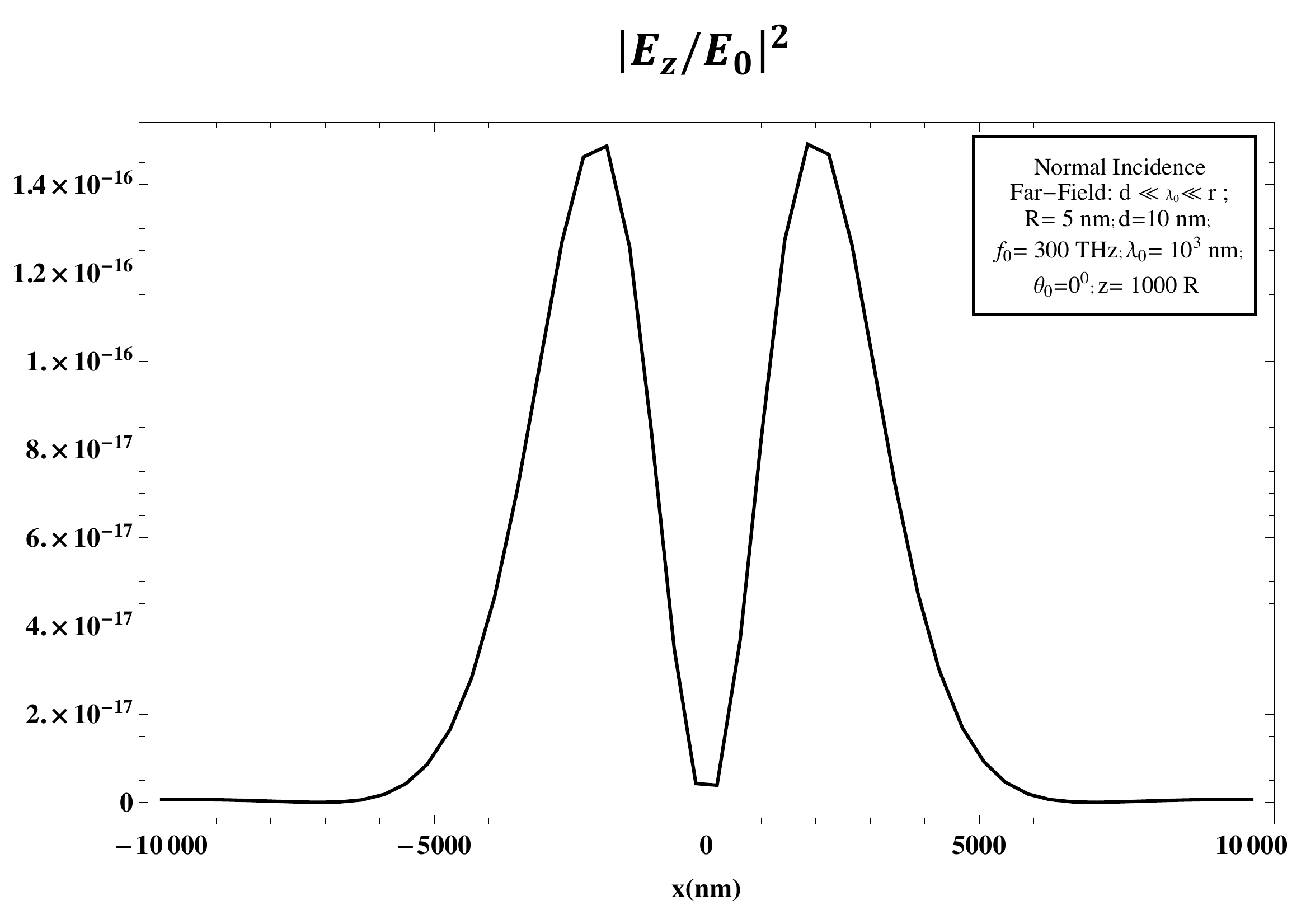}
(c)\includegraphics[width=8.0cm,height=6.0cm]{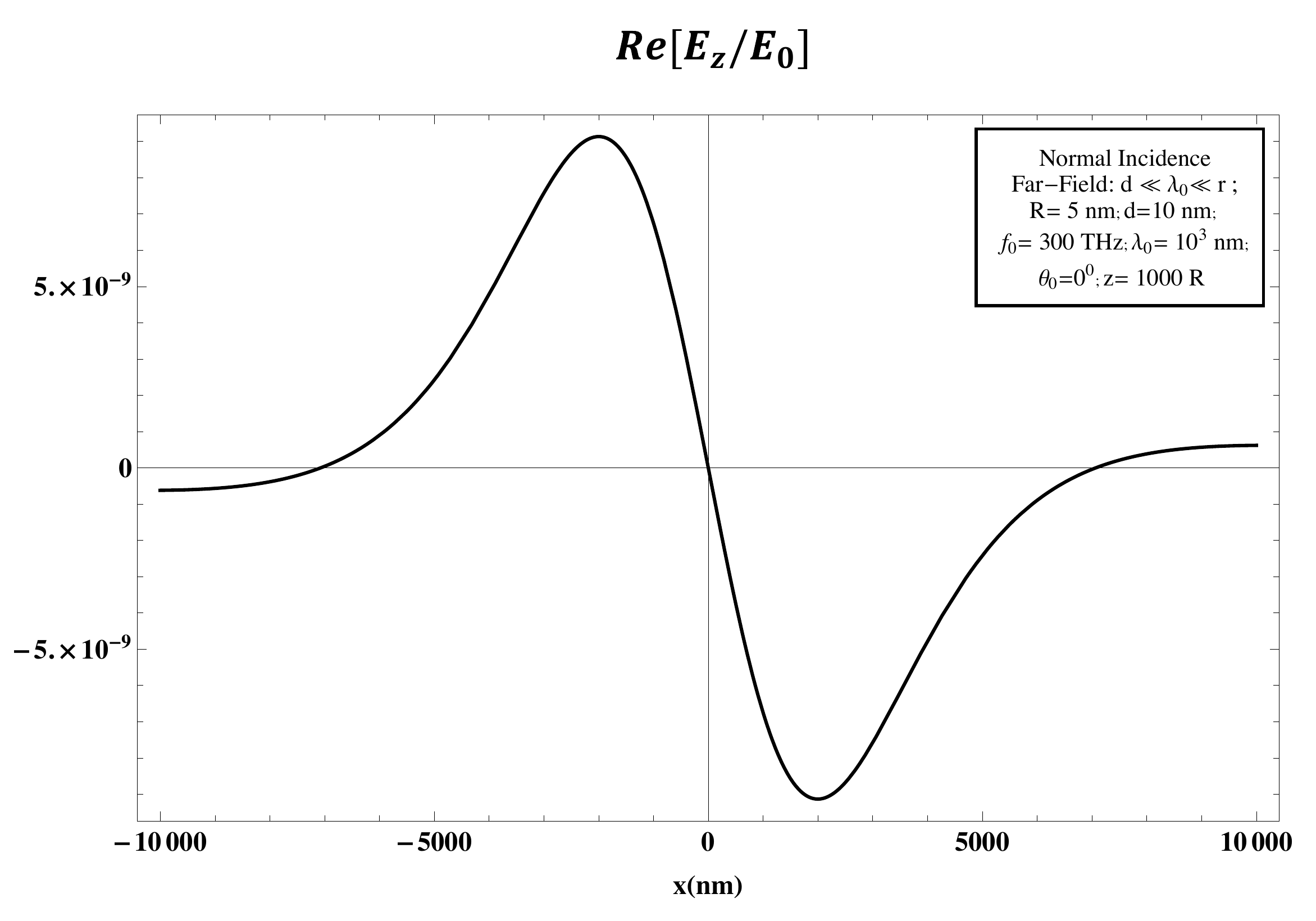}
\caption{ $\mid {E}_{\parallel}(x,y,z;t)/{E}_{0}\mid^{2}$(a),$\mid E_{z}(x,y,z;t)/E_{0}\mid^{2}$(b) and $ Re[E_{x}(x,y,z;t)/E_{0}]$(c) produced by a perforated
2D plasmonic layer of GaAs as a function of lateral distance $ r_{_{\parallel}}= x\,(y=0)$ from the aperture.}
\label{FIGFFT0E}
\end{figure}
\newpage
\subsection{For $ R =5\,nm $, $ \lambda_{0}=1000\,nm $, $ f_{0}=300\,THz $ and $ \theta_{0}=0^{\circ} $}
\subparagraph{For the Near-Field zone: $ z = 50\,R $}.
\begin{figure}[h]
\centering
\includegraphics[width=6cm,height=4.5cm]{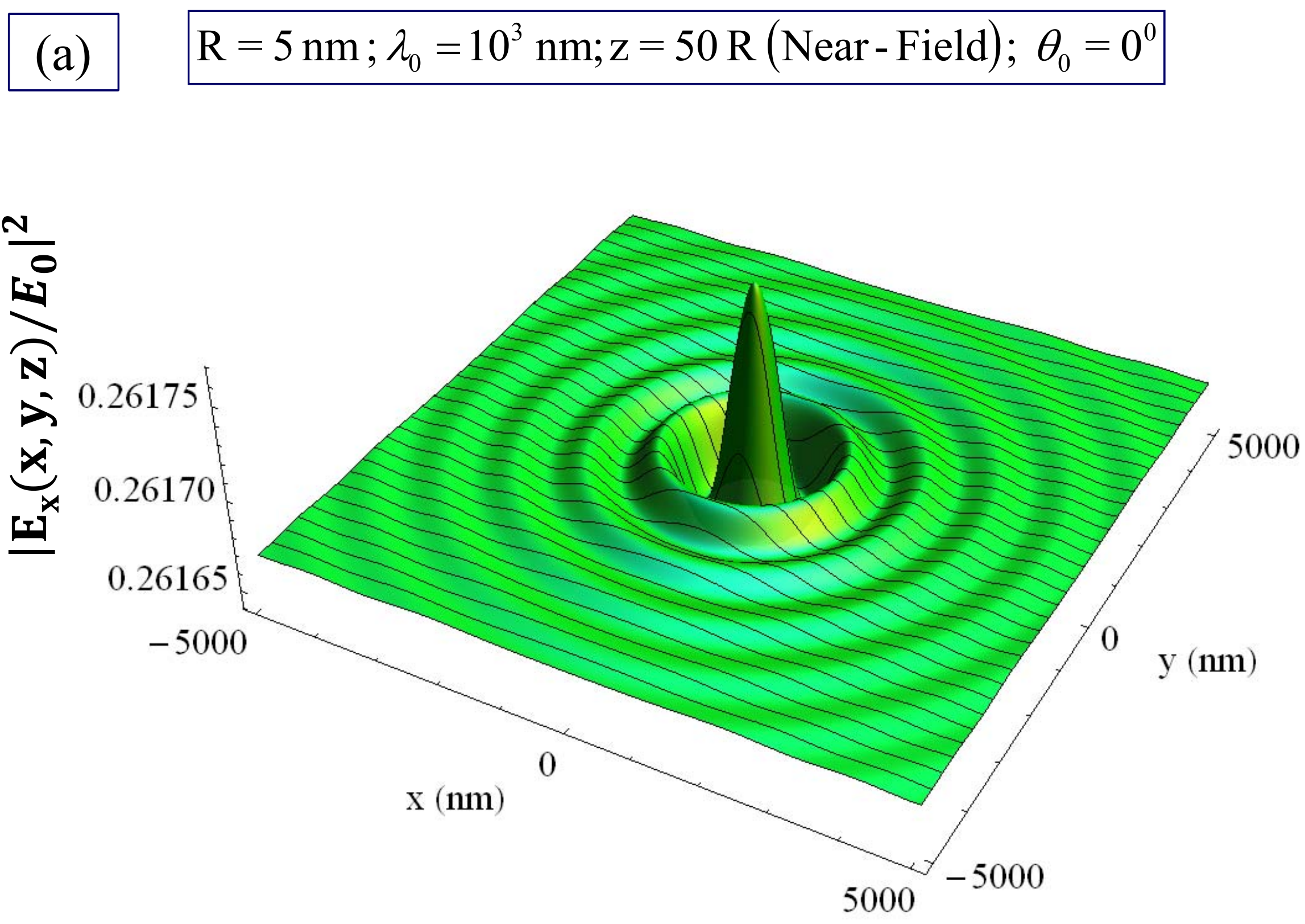}\\
\includegraphics[width=6cm,height=4.750cm]{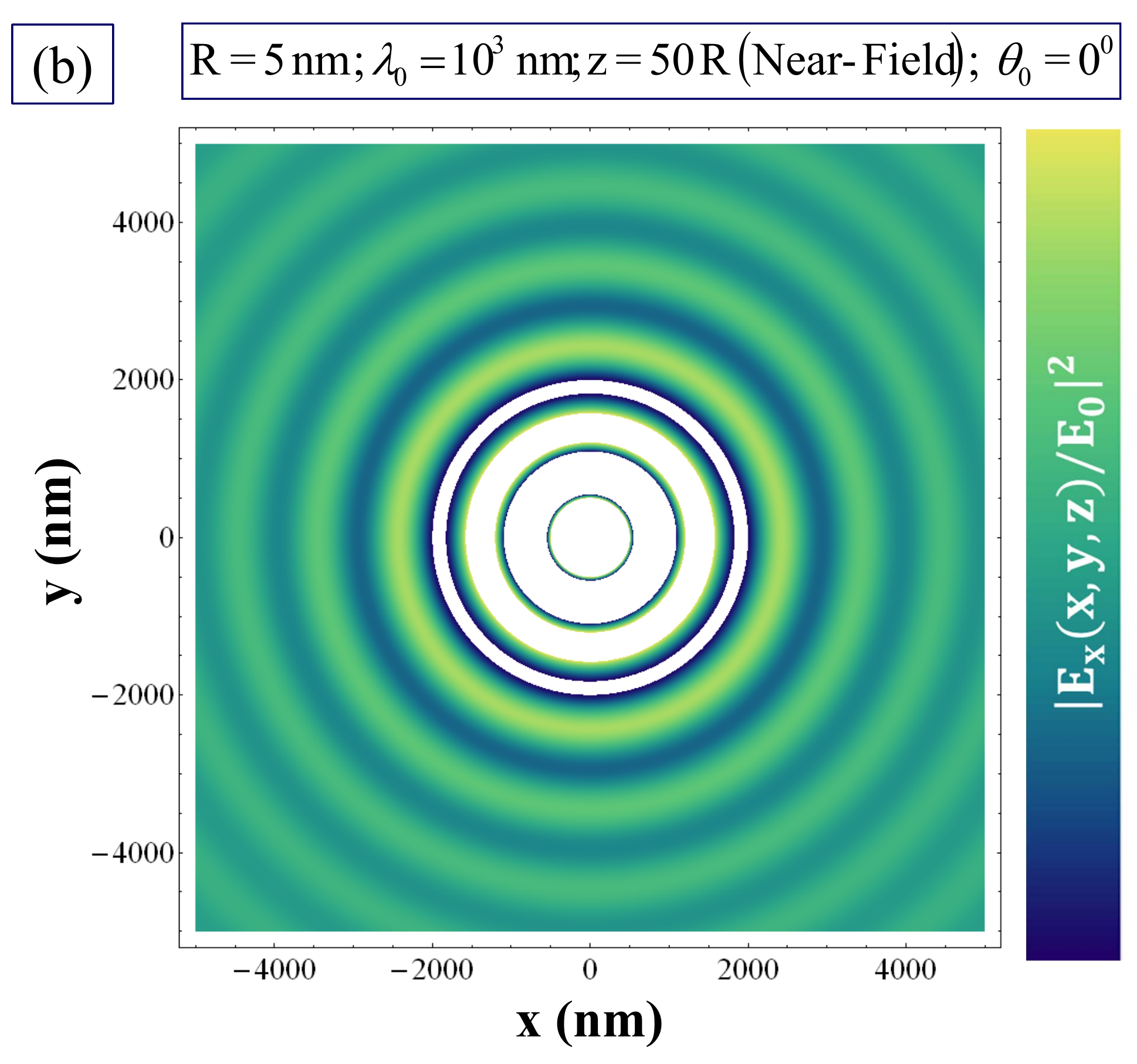}\\
\includegraphics[width=6cm,height=4.5cm]{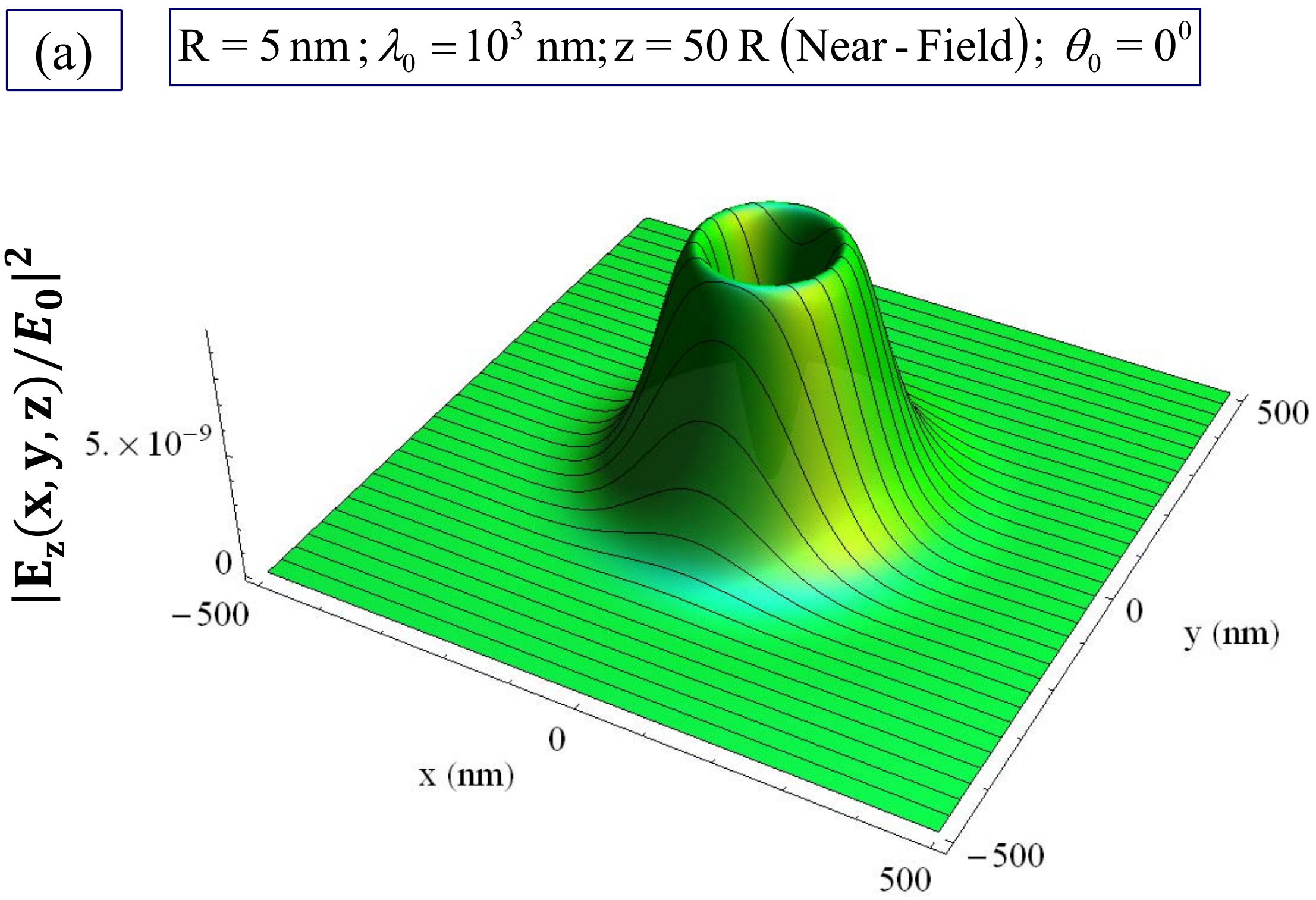}\\
\includegraphics[width=6cm,height=4.75cm]{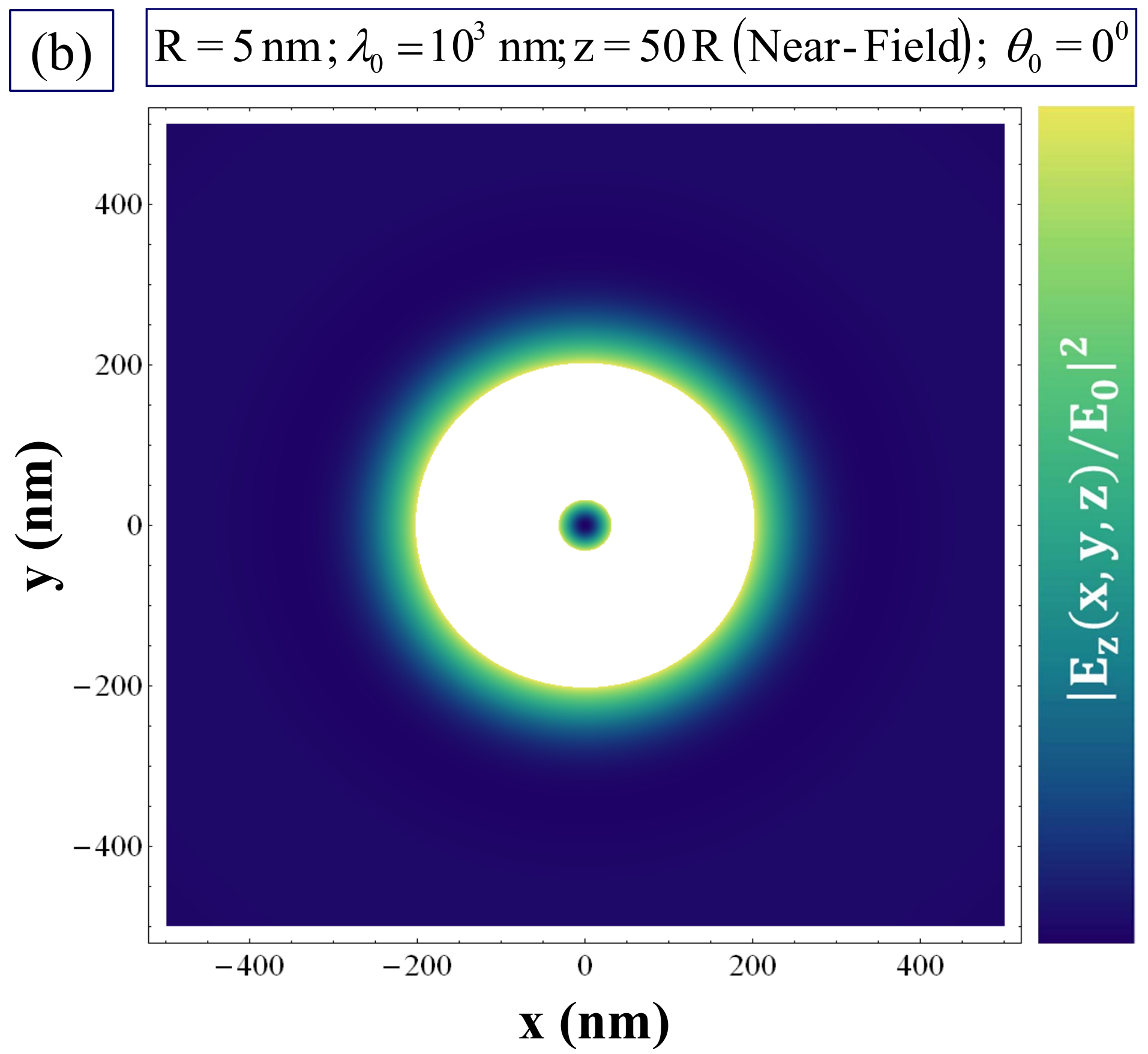}\\
\caption{Field distribution of GaAs layer in terms of 3D $(a's)$ and density $(b's)$ plots:
\newline
$\mid E_{x}(x,y,z;t)/E_{0}\mid^{2}$ and $ \mid E_{z}(x,y,z;t)/E_{0}\mid^{2}$ as functions $ x $ and $ y $ for fixed $ z $.}
\label{FIG3DR5T0N}
\end{figure}
\subparagraph{For the Middle-Field zone: $ z = 300\,R $}.
\begin{figure}[h]
\centering
\includegraphics[width=6cm,height=4.5cm]{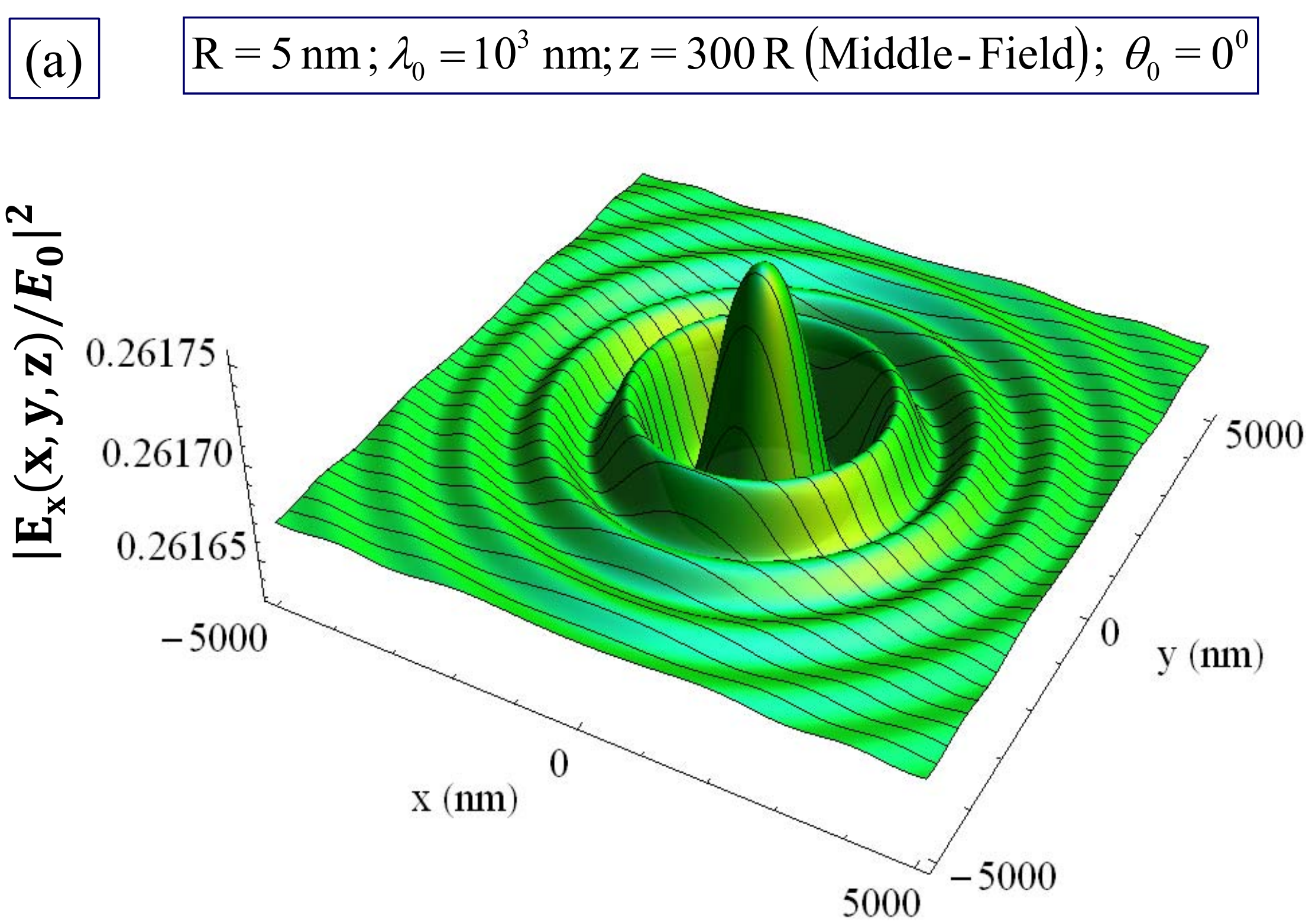}\\
\includegraphics[width=6cm,height=4.75cm]{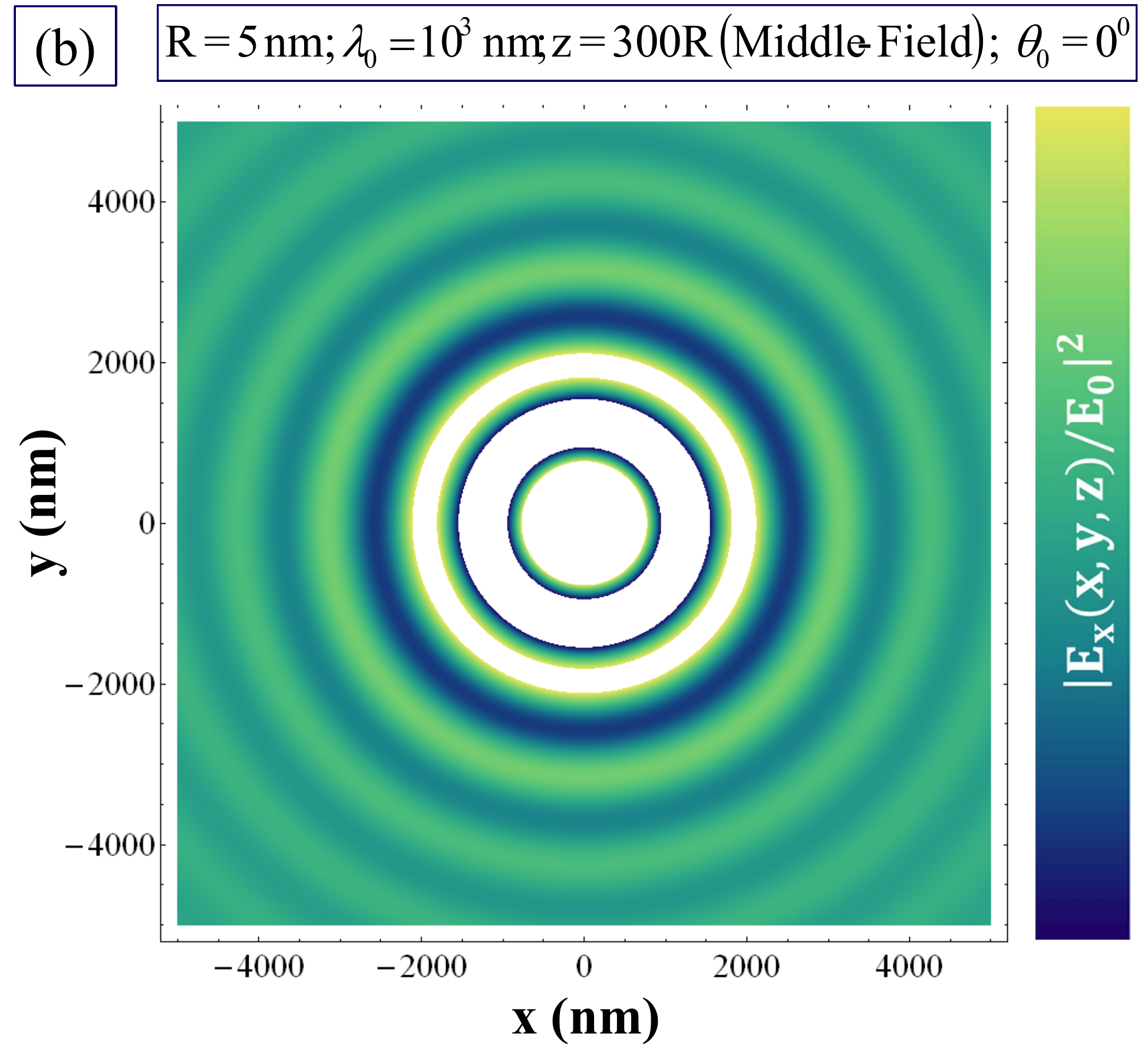}\\
\includegraphics[width=6cm,height=4.5cm]{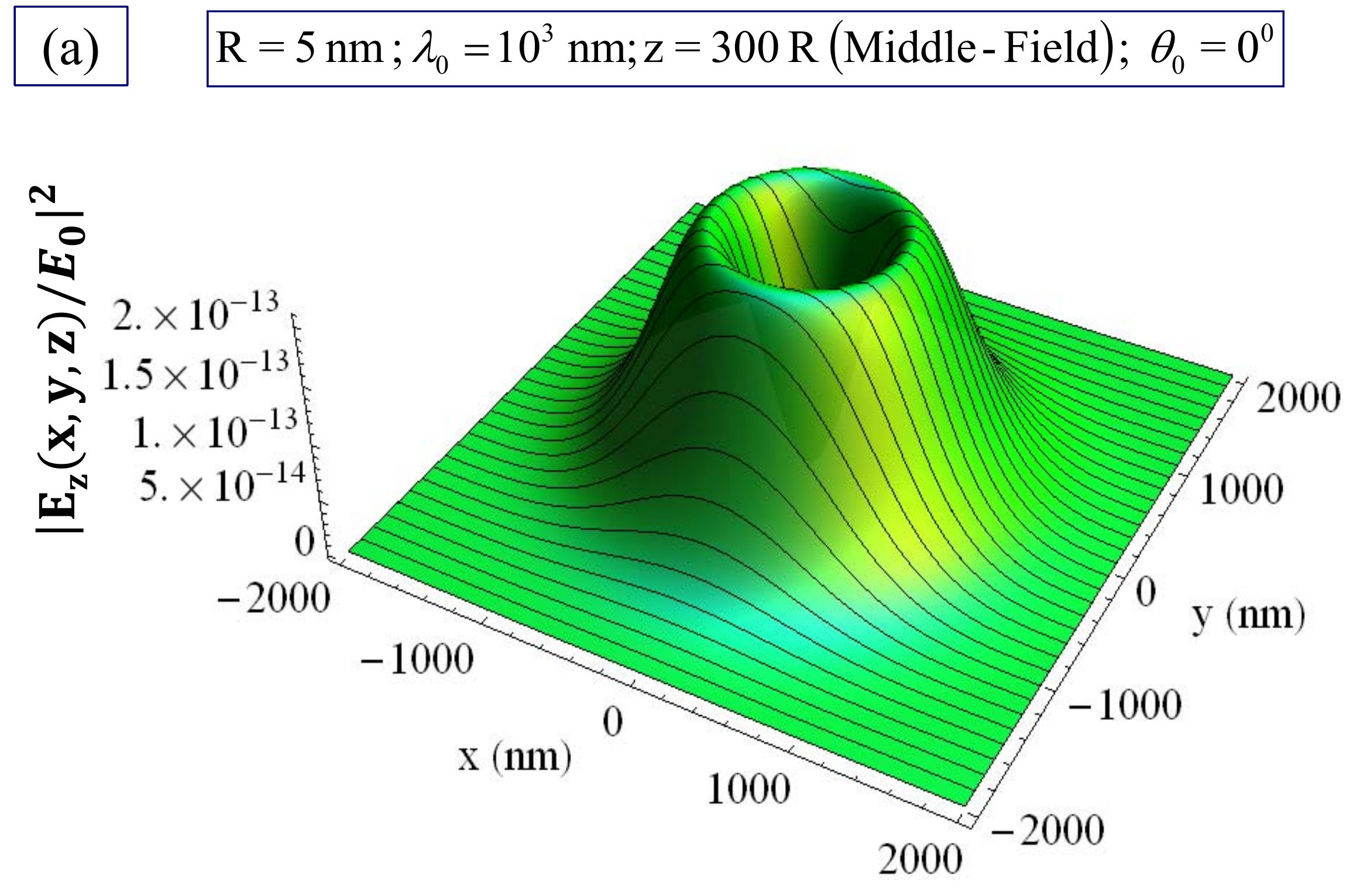}\\
\includegraphics[width=6cm,height=4.75cm]{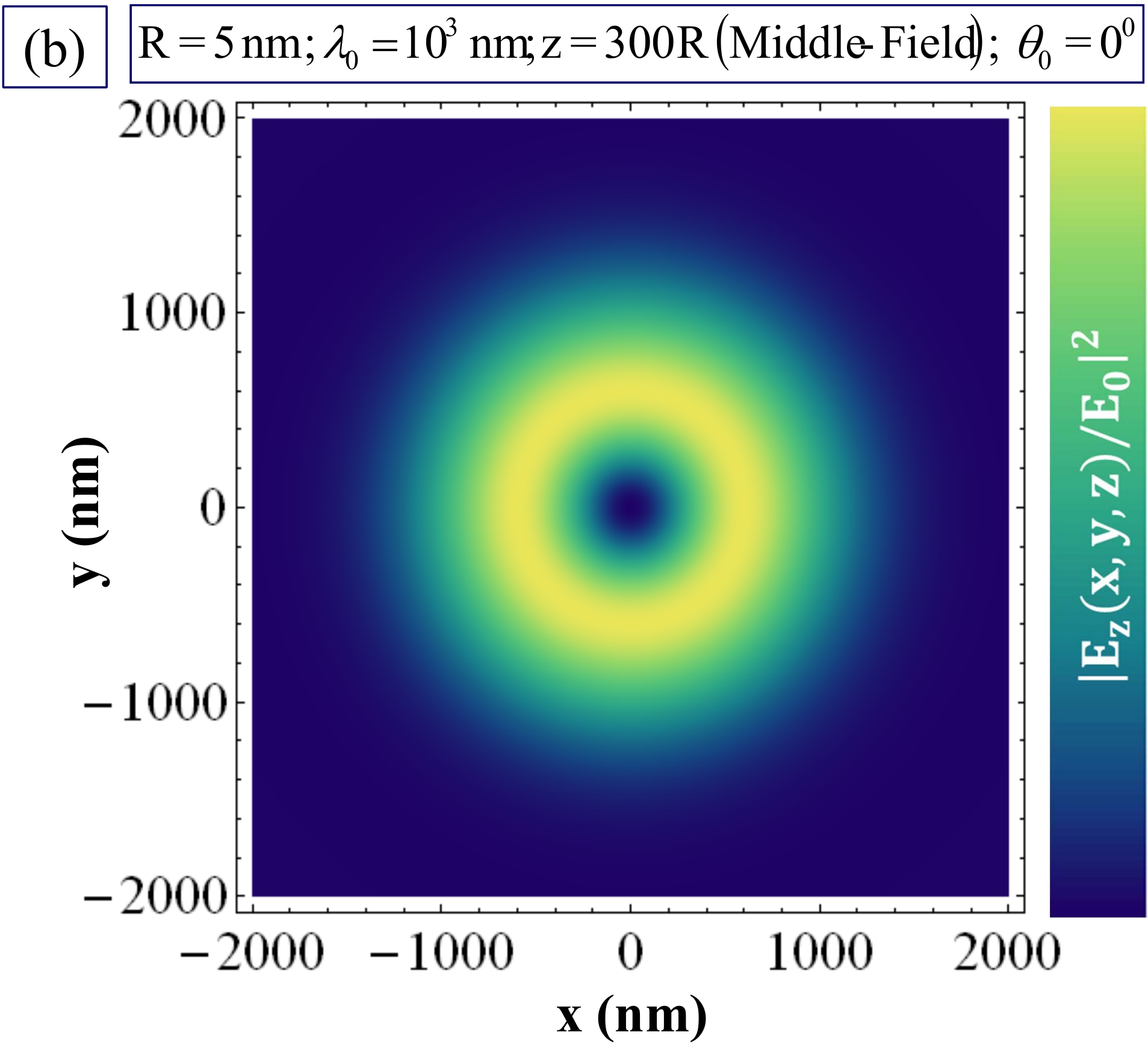}\\
\caption{Field distribution of GaAs layer in terms of 3D $(a's)$ and density $(b's)$ plots:
\newline
$\mid E_{x}(x,y,z;t)/E_{0}\mid^{2}$ and $\mid E_{z}(x,y,z;t)/E_{0}\mid^{2}$ as functions $ x $ and $ y $ for fixed $ z $.}
\label{FIG3DR5T0M}
\end{figure}
\newpage
\subparagraph{For the Far-Field zone: $ z = 1000\,R $}.
\begin{figure}[h]
\centering
\includegraphics[width=6cm,height=4.5cm]{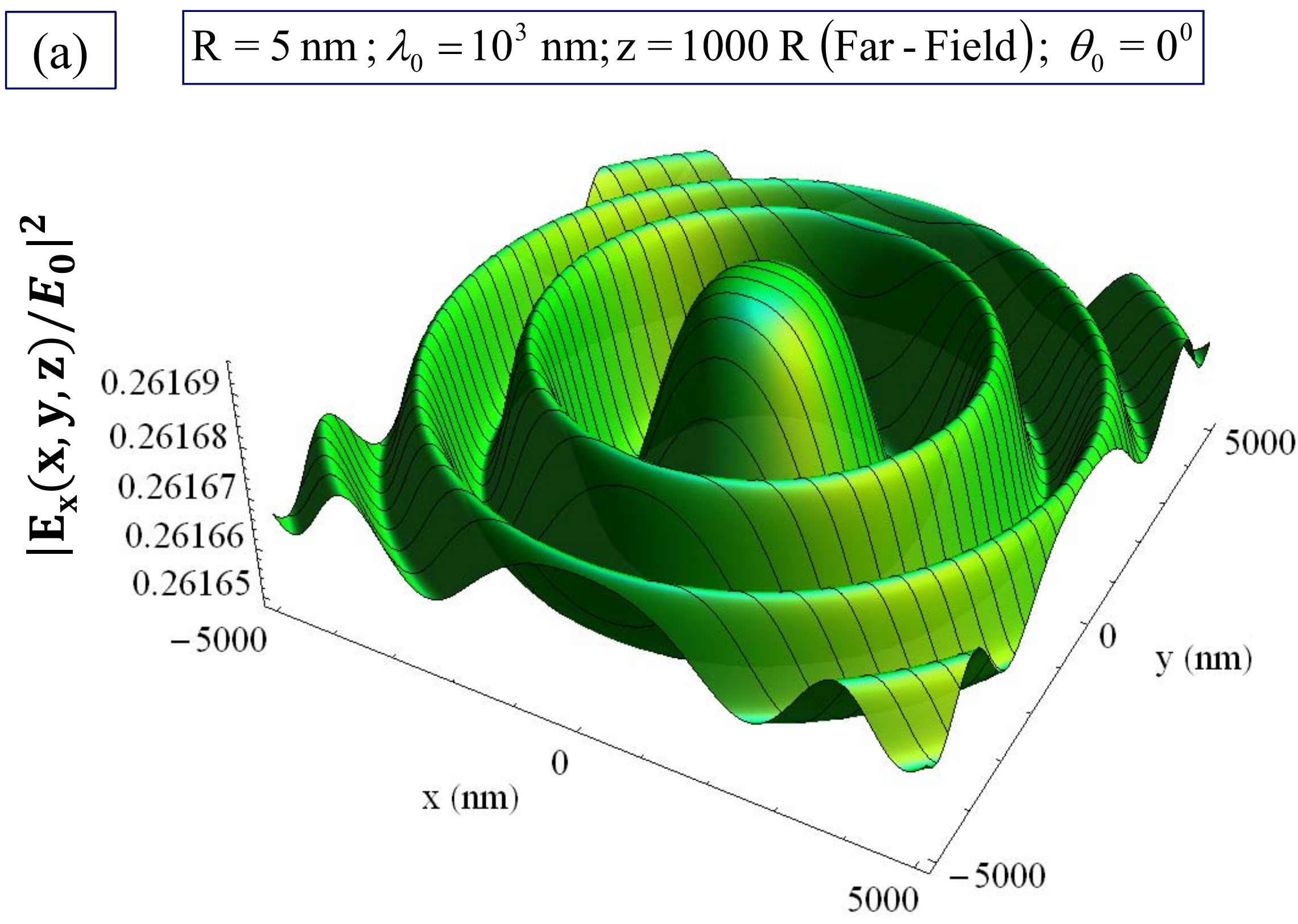}\\
\includegraphics[width=6cm,height=4.75cm]{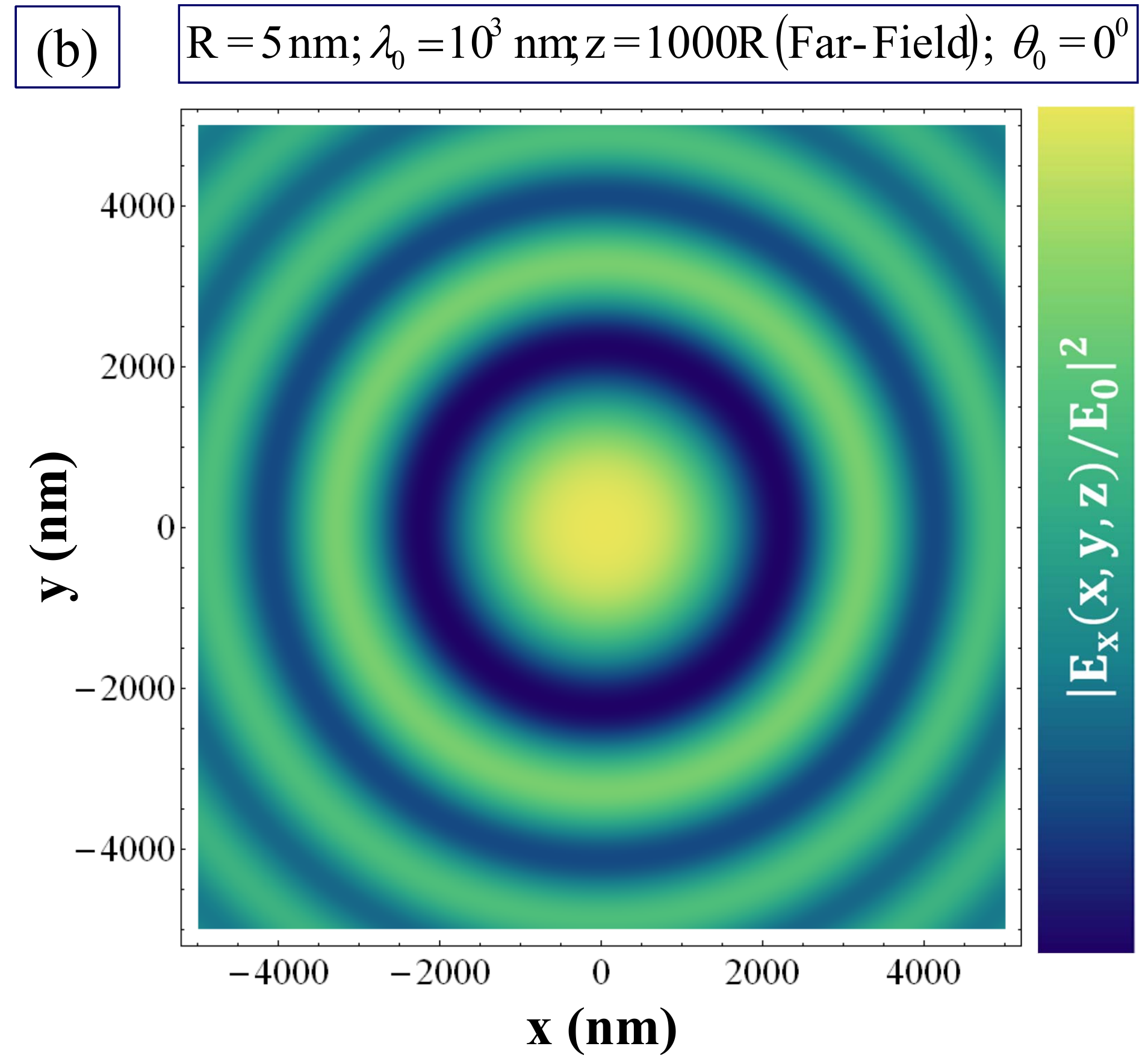}\\
\includegraphics[width=6cm,height=4.50cm]{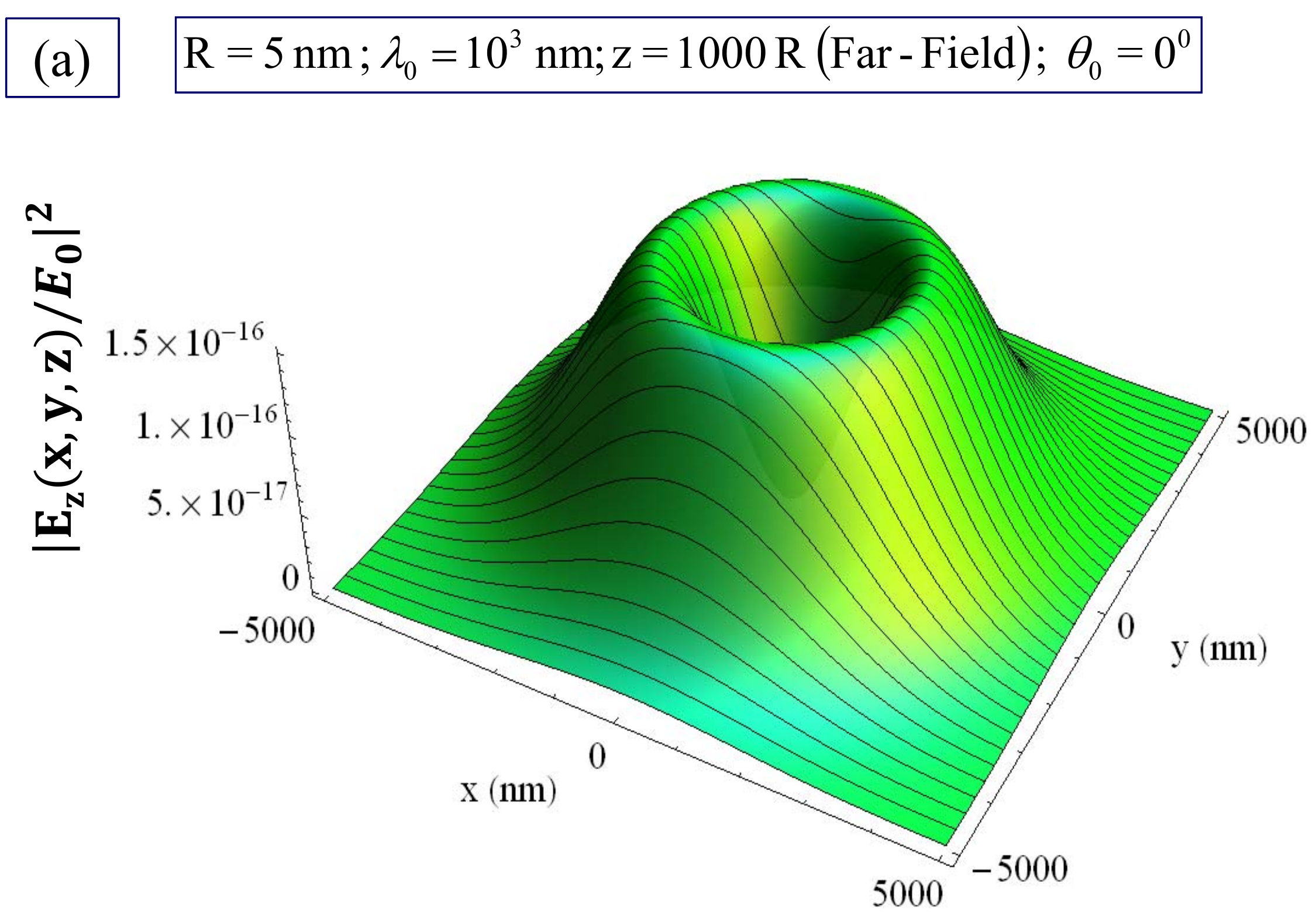}\\
\includegraphics[width=6cm,height=5.0cm]{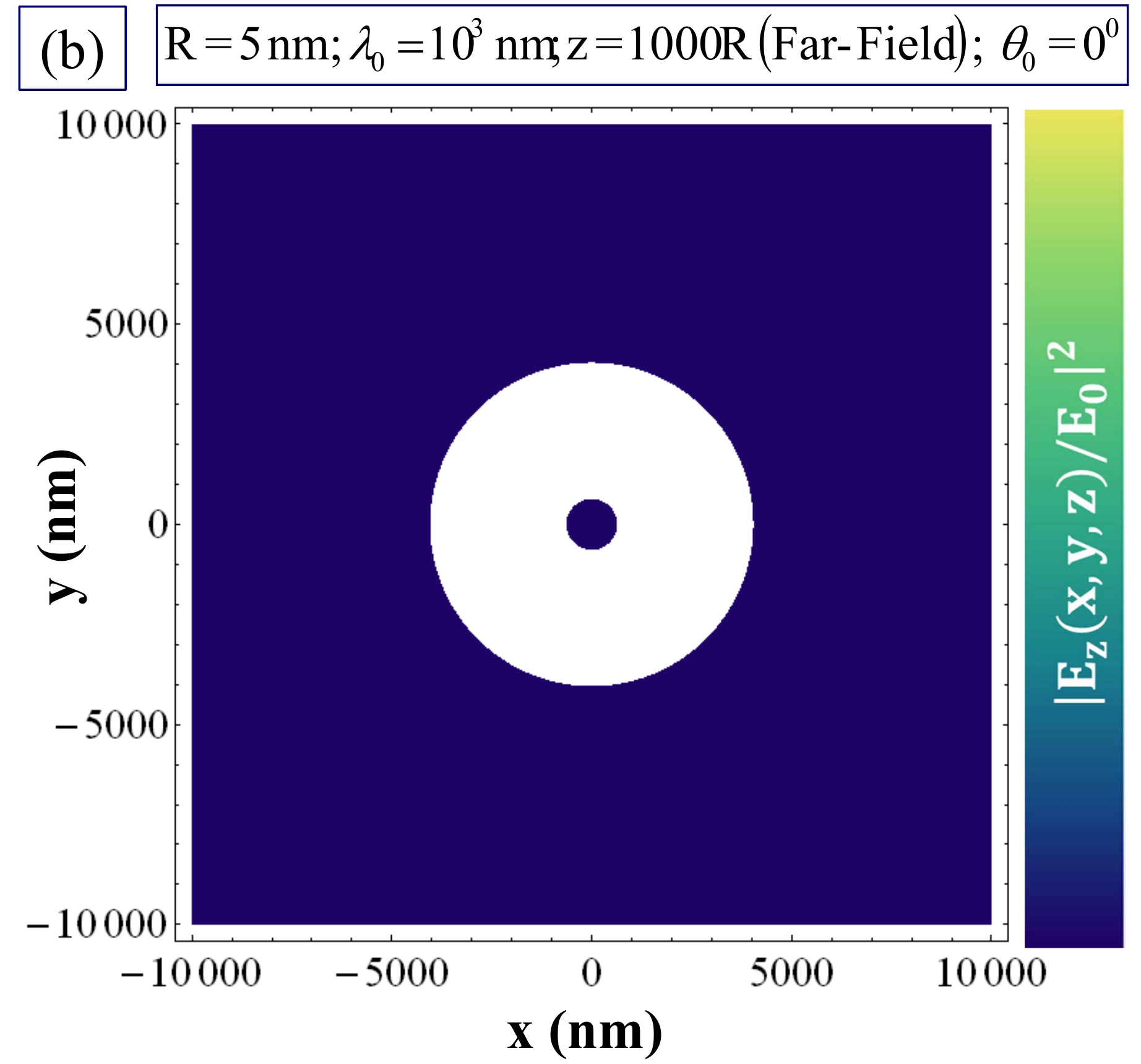}\\
\caption{Field distribution of GaAs layer in terms of 3D $(a's)$ and density $(b's)$ plots:
\newline
$\mid E_{x}(x,y,z;t)/E_{0}\mid^{2}$ and $\mid E_{z}(x,y,z;t)/E_{0}\mid^{2}$ as functions $ x $ and $ y $ for fixed $ z $.}
\label{FIG3DR5T0F}
\end{figure}
\section{Conclusions}
We have examined the transmission of electromagnetic waves through a nano-hole
in a thin plasmonic semiconductor screen for normal incidence of the wave train.
This analysis is based on our previously constructed [4], [5], [6] closed-form
dyadic electromagnetic Green's function for a thin plasmonic/excitonic layer, which
is here adapted to embody a nano-hole (Section 3 above).  The resulting closed-form
dyadic Green's function is employed in the study of electromagnetic wave
transmission through both the hole as well as through the screen itself (Section 4
above).  This approach has facilitated relatively/simple numerical computations exhibited
in Section 5, and is not in any way restricted to a metallic screen.   Moreover, our formulation,
which is based on the use of an \textit{integral} equation for the dyadic Green's function
involved (rather than Maxwell's partial differential equations), automatically incorporates
the boundary conditions, which would otherwise need to be addressed explicitly.
It also incorporates the role of the 2D plasmon of the thin layer, which is smeared by
its lateral wavenumber dependence.

The calculated results shown in the figures of Section 5 exhibit a positional broadening
of the central electromagnetic transmission maximum as $ z$ increases from the near to the
far zones.  This is to be expected as a simple geometric spreading of the radiation
emanating from the nano-hole  itself, but it is also interesting to note that the transmitted
radiation is well supported at large lateral distances from the nano-hole by electromagnetic
transmission \textit{directly} through the plasmonic sheet.  Moreover, the figures also
exhibit interference fringes as a function of lateral distance from the nano-hole in presence
of transmission through the sheet as well as through the hole; and as the lateral distance
from the hole increases this interference diminishes and the transmission tends to flatten
to that emerging through the sheet alone, in the absence of the distant nano-hole.
\appendix
\section{Matrix Elements of $\widehat{\Omega}^{-1}$ and $\widehat{G}_{3D}(\vec{k}_{\parallel},z,0;\omega)$}\label{App:AppendixA}
The elements of $\widehat{\Omega}^{-1}$ are given by(notation: $ a_{0}^{2}=k_{z}^{2}-
\frac{2 i k_{z}}{d}$)
\small
\begin{eqnarray}\label{A.1}
  [\widehat{\Omega}^{-1}]_{_{11}} =  \frac{\left[1+ \left(\frac{\gamma}{2\,i\,k_{z}}
  \right) \left( 1- \frac{k_{y}^{2}}{q_{\omega}^{2}}\right)
 \right]}{\left[\left(1+\left(\frac{\gamma}{2\,i\,k_{z}}\right)\right)\left(1+ \left(\frac{\gamma}{2\,i\,k_{z}}\right) \left( 1- \frac{k_{\parallel}^{2}}{q_{\omega}^{2}}\right) \right)
 \right]},
\end{eqnarray}
\begin{eqnarray}\label{A.2}
  [\widehat{\Omega}^{-1}]_{_{12}} =  \frac{\left[\left(\frac{\gamma}{2\,i\,k_{z}}\right) \left(\frac{k_{x}\,k_{y}}{q_{\omega}^{2}} \right)
 \right]}{\left[\left(1+\left(\frac{\gamma}{2\,i\,k_{z}}\right)\right)\left(1+ \left(\frac{\gamma}{2\,i\,k_{z}}\right) \left( 1- \frac{k_{\parallel}^{2}}{q_{\omega}^{2}}\right) \right)
 \right]},
\end{eqnarray}
\begin{equation}\label{A.3}
    [\widehat{\Omega}^{-1}]_{_{21}} = [\widehat{\Omega}^{-1}]_{_{12}},
 \end{equation}
\begin{equation}\label{A.4}
    [\widehat{\Omega}^{-1}]_{_{31}} = [\widehat{\Omega}^{-1}]_{_{13}}=[\widehat{\Omega}^{-1}]_{_{32}}=[\widehat{\Omega}^{-1}]_{_{23}}=0,
\end{equation}
\begin{eqnarray}\label{A.5}
 [\widehat{\Omega}^{-1}]_{_{22}} = \frac{\left[1+ \left(\frac{\gamma}{2\,i\,k_{z}}\right) \left( 1- \frac{k_{x}^{2}}{q_{\omega}^{2}}\right)
 \right]}{\left[\left(1+\left(\frac{\gamma}{2\,i\,k_{z}}\right)\right)\left(1+ \left(\frac{\gamma}{2\,i\,k_{z}}\right) \left( 1- \frac{k_{\parallel}^{2}}{q_{\omega}^{2}}\right) \right)
 \right]},
\end{eqnarray}
\normalsize
and
\small
\begin{eqnarray}\label{A.6}
  [\widehat{\Omega}^{-1}]_{_{33}} = \frac{1}{\left[1+ \left(\frac{\gamma}{2\,i\,k_{z}}\right) \left( 1- \frac{a_{0}^{2}}{q_{\omega}^{2}}\right)
 \right]}.
\end{eqnarray}
\normalsize
\section{}\label{App:AppendixB}
Furthermore the elements of ${\widehat{G}_{3D}}({\vec{k}}_{\parallel};z,0;
\omega)$ of Eq.(\ref{A2.12}) are given by the matrix ${{G}_{3D}}^{\,ij}({\vec{k}}_{
\parallel};z,0;\omega)$
as:
\begin{equation}\label{B.1}
{G}_{3D}^{xx}\left({\vec{k}}_{\parallel};z,0;\omega\right) =\,-\,\frac{e^{i\,k_{z}\mid z\mid }}{2\,i\,k_{z}}\left(1-\frac{k_{x}^{2}}{q_{\omega}^{2}}\right),
\end{equation}
\begin{equation}\label{B.2}
{G}_{3D}^{yy}\left({\vec{k}}_{\parallel};z,0;\omega\right)=\,-\,\frac{e^{i\,k_{z}\mid z\mid }}{2\,i\,k_{z}}\left(1-\frac{k_{y}^{2}}{q_{\omega}^{2}}\right),
\end{equation}
\begin{equation}\label{B.3}
{G}_{3D}^{zz}\left({\vec{k}}_{\parallel};z,0;\omega\right)=\,-\,\frac{e^{i\,k_{z}\mid z\mid }}{2\,i\,k_{z}}\left(1-\frac{k_{z}^{2}- 2 i k_{z} \delta (z)}{q_{\omega}^{2}}\right),
\end{equation}
\begin{equation}\label{B.4}
{G}_{3D}^{xy}\left({\vec{k}}_{\parallel};z,0;\omega\right)=\,+\,\frac{e^{i\,k_{z}\mid z\mid }}{2\,i\,k_{z}}\left(\frac{k_{x}\,k_{y}}{q_{\omega}^{2}}\right),
\end{equation}
\begin{equation}\label{B.5}
{G}_{3D}^{xz}\left({\vec{k}}_{\parallel};z,0;\omega\right)=\,+\,\frac{e^{i\,k_{z}\mid z\mid }}{2\,i\,k_{z}}\left(\frac{k_{x}\,k_{z}sgn(z)}{q_{\omega}^{2}}\right),
\end{equation}
\begin{equation}\label{B.6}
{G}_{3D}^{yz}\left({\vec{k}}_{\parallel};z,0;\omega\right)=\,+\,\frac{e^{i\,k_{z}\mid z\mid }}{2\,i\,k_{z}}\left(\frac{k_{y}\,k_{z}sgn(z)}{q_{\omega}^{2}}\right),
\end{equation}
\begin{equation}\label{B.7}
{G}_{3D}^{yx}\left({\vec{k}}_{\parallel};z,0;\omega\right) = {G}_{3D}^{xy}\left({\vec{k}}_{\parallel};z,0;\omega\right),
\end{equation}
\begin{equation}\label{B.8}
{G}_{3D}^{xz}\left({\vec{k}}_{\parallel};z,0;\omega\right) = {G}_{3D}^{zx}\left({\vec{k}}_{\parallel};z,0;\omega\right),
\end{equation}
\begin{equation}\label{B.9}
{G}_{3D}^{yz}\left({\vec{k}}_{\parallel};z,0;\omega\right) = {G}_{3D}^{zy}\left({\vec{k}}_{\parallel};z,0;\omega\right).
\end{equation}

Therefore,  Eq.(\ref{A2.16}) may rewritten as
\begin{equation}\label{B.10}
   {\widehat{G}_{fs}}({\vec{k}}_{\parallel};z,0;\omega)={\widehat{G}_{3D}}({
   \vec{k}}_{\parallel};z,0;\omega)\,{\widehat{\Omega}}^{-1}.
\end{equation}
\section{Matrix Elements of $\widehat{G}_{fs}$ $ \longmapsto $ ${G}_{fs}^{ij}(\vec{k}_{\parallel},z,0;\omega)$}\label{App:AppendixC}
\tiny
\begin{eqnarray}\label{C.1}
{G}_{fs}^{xx}\left({\vec{k}}_{\parallel};z,0;\omega\right)&=&\,-\,\frac{e^{i\,k_{z}\mid z\mid }}{2\,i\,k_{z}}
\nonumber\\
&\times&
\left[\frac{1}{D_{1}}\left\{\left(1-\frac{k_{x}^{2}}{q_{\omega}^{2}}\right) +\left(\frac{\gamma}{2\,i\,k_{z}}\right)
\left(1-\frac{k_{\parallel}^{2}}{q_{\omega}^{2}}\right)\right\}\right],
\end{eqnarray}
\begin{eqnarray}\label{C.2}
{G}_{fs}^{yy}\left({\vec{k}}_{\parallel};z,0;\omega\right)&=&\,-\,\frac{e^{i\,k_{z}\mid z\mid }}{2\,i\,k_{z}}
\nonumber\\
&\times&
\left[\frac{1}{D_{1}}\left\{\left(1-\frac{k_{y}^{2}}{q_{\omega}^{2}}\right) +\left(\frac{\gamma}{2\,i\,k_{z}}\right)
\left(1-\frac{k_{\parallel}^{2}}{q_{\omega}^{2}}\right)\right\}\right],
\end{eqnarray}
\normalsize
\small
\begin{equation}\label{C.3}
{G}_{fs}^{zz}\left({\vec{k}}_{\parallel};z,0;\omega\right)=\,-\,\frac{e^{i\,k_{z}\mid z\mid }}{2\,i\,k_{z}}\left[\frac{1}{D_{2}}\left\{\left(1-\frac{k_{z}^{2}- 2 i k_{z} \delta (z)}{q_{\omega}^{2}}\right)\right\}\right],
\end{equation}
\begin{equation}\label{C.4}
{G}_{fs}^{xy}\left({\vec{k}}_{\parallel};z,0;\omega\right)=\,+\,\frac{e^{i\,k_{z}\mid z\mid }}{2\,i\,k_{z}}\left[\frac{1}{D_{1}}\left(\frac{k_{x}\,k_{y}}{q_{\omega}^{2}}\right)\right],
\end{equation}
\begin{equation}\label{C.5}
{G}_{fs}^{xz}\left({\vec{k}}_{\parallel};z,0;\omega\right)
=\,+\,\frac{e^{i\,k_{z}\mid z\mid }}{2\,i\,k_{z}}\left[\frac{1}{D_{2}}\left(\frac{k_{x}\,k_{z}sgn(z)}{q_{\omega}^{2}}\right)\right],
\end{equation}
\begin{equation}\label{C.6}
{G}_{fs}^{yz}\left({\vec{k}}_{\parallel};z,0;\omega\right)=\,+\,\frac{e^{i\,k_{z}\mid z\mid }}{2\,i\,k_{z}}\left[\frac{1}{D_{2}}\left(\frac{k_{y}\,k_{z}sgn(z)}{q_{\omega}^{2}}\right)\right],
\end{equation}
\begin{equation}\label{C.7}
{G}_{fs}^{yx}\left({\vec{k}}_{\parallel};z,0;\omega\right) = {G}_{fs}^{xy}\left({\vec{k}}_{\parallel};z,0;\omega\right),
\end{equation}
\begin{eqnarray}\label{C.8}
{G}_{fs}^{zx}\left({\vec{k}}_{\parallel};z,0;\omega\right)&=&\frac{e^{i\,k_{z}\mid z\mid }}{2\,i\,k_{z}}
\nonumber\\
&\times&
\left[\frac{1}{D_{1}}\left(\left(1+\left(\frac{\gamma}{2
\,i\,k_{z}}\right)\right)\frac{k_{z}\,k_{x}sgn(z)}{q_{\omega}^{2}}\right)\right],
\nonumber\\
\end{eqnarray}
\begin{eqnarray}\label{C.9}
{G}_{fs}^{zy}\left({\vec{k}}_{\parallel};z,0;\omega\right)
&=&\frac{e^{i\,k_{z}\mid z\mid }}{2\,i\,k_{z}}
\nonumber\\
&\times&
\left[\frac{1}{D_{1}}\left(\left(1+\left(\frac{
\gamma}{2\,i\,k_{z}}\right)\right)\frac{k_{z}\,k_{y}sgn(z)}{q_{\omega}^{2}}\right)\right],
\nonumber\\
\end{eqnarray}
\begin{equation}\label{C.10}
{G}_{fs}^{xz}\left({\vec{k}}_{\parallel};z,0;\omega\right) = {G}_{fs}^{zx}\left({\vec{k}}_{\parallel};z,0;\omega\right),
\end{equation}
and
\begin{equation}\label{C.11}
   D_{1}=\left(1+\left(\frac{\gamma}{2\,i\,k_{z}}\right)\right)\left[1+ \left(\frac{\gamma}{2\,i\,k_{z}}\right) \left( 1- \frac{k_{\parallel}^{2}}{q_{\omega}^{2}}\right) \right],
\end{equation}
\begin{equation}\label{C.12}
   D_{2}=\left[1+ \left(\frac{\gamma}{2\,i\,k_{z}}\right) \left( 1-
   \frac{a_{0}^{2}}{q_{\omega}^{2}}\right) \right].
\end{equation}
\normalsize
It should be noted that, in the text above, we choose the coordinate system such that
$ k_{y}\equiv 0 $.  In this case, $ k_{\parallel}=k_{x} $ and $ k_{z}^{2}=q_{\omega}^{2}-
k_{x}^{2}$.  Moreover, with $ k_{y}=0 $, $
{\widehat{G}_{3D}}({\vec{k}}_{\parallel};0,0;\omega)$ becomes diagonal.
\section{Matrix Elements of $\widehat{T}^{0}$ }\label{App:AppendixD}
Since $ \widehat{\overline{\overline{G}}}_{fs}(\vec{k}_{0_{\parallel}};0,0;
\omega_{0})$ is diagonal, we have
\begin{eqnarray}\label{D.1}
  {\widehat{T}^{0}}  & = &
  \begin{bmatrix}
  {G}^{xx} & 0 & {G}^{xz} \\
  0 & {G}^{yy} & 0 \\
  {G}^{zx} & 0 & {G}^{zz}
 \end{bmatrix}.
\nonumber\\
&\times&
\begin{bmatrix}
  1+\gamma_{0}{\overline{\overline{G}}}_{fs}^{\,xx} & 0 & 0 \\
  0 & 1+\gamma_{0}{\overline{\overline{G}}}_{fs}^{\,yy} & 0 \\
  0 & 0 & 1+\gamma_{0}{\overline{\overline{G}}}_{fs}^{\,zz}
 \end{bmatrix}
\end{eqnarray}
or
\begin{eqnarray}\label{D.2}
  {\widehat{T}^{0}}  & = &
\begin{bmatrix}
  {T}_{xx}^{0} & 0 & {T}_{xz}^{0} \\
  0 & {T}_{yy}^{0} & 0 \\
  {T}_{zx}^{0} & 0 & {T}_{zz}^{0}
 \end{bmatrix}
\end{eqnarray}
where the matrix elements are given by
\begin{equation}\label{D.3}
\left\{
         \begin{array}{ll}
            {T}_{xx}^{0} = & \hbox{$ G^{xx}\left[ 1\,+\,\gamma_{0}\,\overline{
            \overline{G}}_{fs}^{\,xx}\right]$ ; (a)} \\
             {T}_{zx}^{0} = & \hbox{$ G^{zx}\left[ 1\,+\,\gamma_{0}\,\overline{
             \overline{G}}_{fs}^{\,xx}\right]$ ; (b)} \\
            {T}_{yy}^{0} = & \hbox{$ G^{yy}\left[ 1\,+\,\gamma_{0}\,\overline{
            \overline{G}}_{fs}^{\,yy}\right]$ ; (c)} \\
             {T}_{xz}^{0} = & \hbox{$ G^{xz}\left[ 1\,+\,\gamma_{0}\,\overline{
             \overline{G}}_{fs}^{\,zz}\right]$ ; (d)} \\
            {T}_{zz}^{0} = & \hbox{$ G^{zz}\left[ 1\,+\,\gamma_{0}\,\overline{
            \overline{G}}_{fs}^{\,zz}\right]$ . (e)}
         \end{array}
       \right.
\end{equation}
In order to facilitate the calculations of the above matrix elements, it is important to note that
$ \widehat{G}=\widehat{G}(\vec{r}_{\parallel},0;z,0;\omega_{0}) $
and $ \widehat{\overline{\overline{G}}}_{fs}= \widehat{\overline{\overline{G}}}_{fs}(\vec{k}_{0_{\parallel}};0,0;\omega_{0})$ are
 evaluated at the incident wave frequency $
\omega_{0}$.  Therefore, the matrix elements of the dyad $ {\widehat{T}^{0}} $
are given by (note that $\theta_{0}$ is the angle of incident)
\begin{equation}\label{D.4}
    {T}_{xx}^{0} = G_{fs}^{xx}\left[\frac{1\,+\,\gamma_{0}\,\overline{
    \overline{G}}_{fs}^{\,xx}}{1\,
    +\,\beta_{0}\,{G}_{fs0}^{\,xx}}\right] \,\,\,\,
\end{equation}
where
\begin{equation}\label{D.4I}
    \left\{
       \begin{array}{ll}
         \overline{\overline{G}}_{fs}^{\,xx} \equiv\, & \hbox{$\frac{\,-\,\cos(\theta_{0})}{2\,i\,\left[q_{\omega_{0}}\,+\,\Gamma_{0}\,\cos(\theta_{0})\right]}$;    (a)} \\
         {G}_{fs0}^{\,xx} \equiv\, & \hbox{$ {G}_{fs0}^{\,xx}(0-0;0,0;\omega_{0}) $;     (b)} \\
         {G}_{fs}^{\,xx} \equiv\, & \hbox{$ {G}_{fs}^{\,xx}(\vec{r}_{\parallel},0;z,0;\omega_{0})$.    (c)}
       \end{array}
     \right.
\end{equation}
\begin{equation}\label{D.5}
    {T}_{yy}^{0} = G_{fs}^{yy}\left[\frac{1\,+\,\gamma_{0}\,\overline{\overline{G}}_{fs}^{\,yy}}{1\,+\,\beta_{0}\,{G}_{fs0}^{\,yy}}\right] \,\,\,\,
\end{equation}
where
\begin{equation}\label{D.5I}
    \left\{
       \begin{array}{ll}
         \overline{\overline{G}}_{fs}^{\,yy} \equiv\, & \hbox{$\frac{\,-\,1}{2\,i\,\left[\,\Gamma_{0}\,+\,q_{\omega_{0}}\,\cos(\theta_{0})
         \right]}$;    (a)} \\
         {G}_{fs0}^{\,yy} \equiv\, & \hbox{$ {G}_{fs0}^{\,yy}(0-0;0,0;\omega_{0}) $;     (b)} \\
         {G}_{fs}^{\,yy} \equiv\, & \hbox{$ {G}_{fs}^{\,yy}(
         \vec{r}_{\parallel},0;z,0;\omega_{0})$.    (c)}
       \end{array}
     \right.
\end{equation}
\begin{equation}\label{D.6}
    {T}_{zz}^{0} = G_{fs}^{zz}\left[\frac{1\,+\,\gamma_{0}\,\overline{\overline{G}}_{fs}^{\,zz}}{1\,+\,\beta_{0}\,{G}_{fs0}^{\,zz}}\right] \,\,\,\,
\end{equation}
where
\small
\begin{equation}\label{D.6I}
    \left\{
       \begin{array}{ll}
         \overline{\overline{G}}_{fs}^{\,zz} \equiv\, & \hbox{$  \frac{\,-\,1}{2\,i} \left\{\frac{q_{\omega_{0}}\,\sin^{2}(\theta_{0})+2\,i\,\delta(0)\cos(\theta_{0})}{\cos(\theta_{0})\,
\left[\,q_{\omega_{0}}^{2}\,+\,2\,i\,\delta(0)\Gamma_{0}\right]\,+\,\Gamma_{0}\,q_{\omega_{0}}\,\sin^{2}(\theta_{0})}\right\}$;    (a)} \\
         {G}_{fs0}^{\,zz} \equiv\, & \hbox{$ {G}_{fs0}^{\,zz}(0-0;0,0;\omega_{0}) $;     (b)} \\
         {G}_{fs}^{\,zz}\equiv\, & \hbox{$ {G}_{fs}^{\,zz}(\vec{r}_{\parallel},0;z,0;\omega_{0})$.    (c)}
       \end{array}
     \right.
\end{equation}
\normalsize
\begin{equation}\label{D.7}
    {T}_{xz}^{0} = G_{fs}^{xz}\left[\frac{1\,+\,\gamma_{0}\,\overline{\overline{G}}_{fs}^{\,zz}}{1\,+\,\beta_{0}\,{G}_{fs0}^{\,zz}}\right] \,\,\,\,
\end{equation}
where
\small
\begin{equation}\label{D.7I}
    \left\{
       \begin{array}{ll}
         \overline{\overline{G}}_{fs}^{\,zz} \equiv\, & \hbox{$  \frac{\,-\,1}{2\,i} \left\{\frac{q_{\omega_{0}}\,\sin^{2}(\theta_{0})\,+2\,i\,\delta(0)\cos(\theta_{0})}{\cos(\theta_{0})\,
\left[\,q_{\omega_{0}}^{2}\,+\,2\,i\,\delta(0)\,\Gamma_{0}\right]\,+\,\Gamma_{0}\,q_{\omega_{0}}\,\sin^{2}(\theta_{0})}\right\}$;    (a)} \\
         {G}_{fs0}^{\,zz} \equiv\, & \hbox{$ {G}_{fs0}^{\,zz}(0-0;0,0;\omega_{0}) $;     (b)} \\
         {G}_{fs}^{\,xz} \equiv\, & \hbox{$ {G}_{fs}^{\,xz}(\vec{r}_{\parallel},0;z,0;\omega_{0})$.    (c)}
       \end{array}
     \right.
\end{equation}
\normalsize
and
\begin{equation}\label{D.8}
    {T}_{zx}^{0} = G_{fs}^{zx}\left[\frac{1\,+\,\gamma_{0}\,\overline{\overline{G}}_{fs}^{\,xx}}{1\,+\,\beta_{0}\,{G}_{fs0}^{\,xx}}\right] \,\,\,\,
\end{equation}
where
\begin{equation}\label{D.8I}
    \left\{
       \begin{array}{ll}
         \overline{\overline{G}}_{fs}^{\,xx} \equiv\, & \hbox{$\frac{\,-\,\cos(\theta_{0})}{2\,i\,\left[q_{\omega_{0}}\,+\,\Gamma_{0}\,\cos(\theta_{0})\right]}$;    (a)} \\
         {G}_{fs0}^{\,xx} \equiv\, & \hbox{$ {G}_{fs0}^{\,xx}(0-0;0,0;\omega_{0}) $;     (b)} \\
         {G}_{fs}^{\,zx} \equiv\, & \hbox{$ {G}_{fs}^{\,zx}(\vec{r}_{\parallel},0;z,0;\omega_{0})$.    (c)}
       \end{array}
     \right.
\end{equation}
\begin{acknowledgments}
This work was partially supported by the NSF-AGEP program.
GG was supported by  contract \# FA 9453-13-1-0291 of AFRL.
\end{acknowledgments}


\end{document}